\documentclass[revtex4]{emulateapj}
\usepackage{amssymb,natbib,graphicx,epstopdf,color,lineno}
\usepackage{lscape}

\def\eg{{e.g.,~}}
\def\ie{{i.e.,~}}

\begin{document}
\label{firstpage}

\title{Detection of the cosmic $\gamma$-ray horizon from multiwavelength observations of blazars}

\author{{\sc A. Dom\'inguez}\altaffilmark{1}, {\sc J.~D. Finke}\altaffilmark{2}, {\sc F. Prada}\altaffilmark{3,4,5}, {\sc J.~R. Primack}\altaffilmark{6}, {\sc F.~S. Kitaura}\altaffilmark{7}, {\sc B. Siana}\altaffilmark{1}, {\sc D. Paneque}\altaffilmark{8,9}}

\slugcomment{Draft; \today}

\shorttitle{Detection of the cosmic $\gamma$-ray horizon}
\shortauthors{DOM\'INGUEZ ET AL.}

\altaffiltext{1}{Department of Physics \& Astronomy, University of California, Riverside, CA 92521, USA; albertod@ucr.edu}
\altaffiltext{2}{U.S. Naval Research Laboratory, Space Science Division, Code 7653, 4555 Overlook Avenue SW, Washington, DC, 20375 USA}
\altaffiltext{3}{Campus of International Excellence UAM+CSIC, Cantoblanco, E-28049 Madrid, Spain}
\altaffiltext{4}{Instituto de F\'{\i}sica Te\'orica, (UAM/CSIC), Universidad Aut\'onoma de Madrid, Cantoblanco, E-28049 Madrid, Spain}
\altaffiltext{5}{Instituto de Astrof\'{\i}sica de Andaluc\'{\i}a (CSIC), Glorieta de la Astronom\'{\i}a, E-18080 Granada, Spain}
\altaffiltext{6}{Department of Physics, University of California, Santa Cruz, CA 95064, USA}
\altaffiltext{7}{Leibniz-Institut f\"ur Astrophysik (AIP), An der Sternwarte 16, D-14482 Potsdam, Germany}
\altaffiltext{8}{Kavli Institute for Particle Astrophysics and Cosmology, SLAC, Stanford University, Stanford, CA 94305, USA}
\altaffiltext{9}{Max-Planck-Institut f\"ur Physik, D-80805 M\"unchen, Germany}



\begin{abstract}
The first statistically significant detection of the cosmic $\gamma$-ray horizon (CGRH) that is independent of any extragalactic background light (EBL) model is presented. The CGRH is a fundamental quantity in cosmology. It gives an estimate of the opacity of the Universe to very high energy (VHE) $\gamma$-ray photons due to photon-photon pair production with the EBL. The only estimations of the CGRH to date are predictions from EBL models and lower limits from $\gamma$-ray observations of cosmological blazars and $\gamma$-ray bursts. Here, we present homogeneous synchrotron/synchrotron self-Compton (SSC) models of the spectral energy distributions of 15 blazars based on (almost) simultaneous observations from radio up to the highest energy $\gamma$-rays taken with the \emph{Fermi} satellite. These synchrotron/SSC models predict the unattenuated VHE fluxes, which are compared with the observations by imaging atmospheric Cherenkov telescopes. This comparison provides an estimate of the optical depth of the EBL, which allows a derivation of the CGRH through a maximum likelihood analysis that is EBL-model independent. We find that the observed CGRH is compatible with the current knowledge of the EBL.

\end{abstract}


\section{Introduction}
\label{sec:intro}

Very high energy (VHE; 30 GeV--30 TeV) photons do not travel unimpeded through cosmological distances in the Universe. A flux attenuation is expected due to photon-photon pair production with the lower energy photons of the extragalactic background light (EBL) in the ultraviolet, optical, and infrared (IR; \eg \citealt{nikishov62,gould67,stecker92,salamon98}). The EBL is the radiation emitted by star formation processes (star light and star light absorbed/re-emitted by dust) plus a small contribution from active galactic nuclei (AGNs) integrated over redshift over all the cosmic star-formation history (\eg \citealt{hauser01,dwek12}). Due to the properties of the interaction, a VHE photon of a given energy interacts mainly with an EBL photon within a well defined and narrow wavelength range. Therefore, a signature of the EBL spectral distribution is expected in the observed VHE spectra of extragalactic sources (\citealt{ackermann12b,abramowski13,yuan13}).


An interesting feature in the observed VHE spectra of extragalactic sources as a consequence of EBL absorption is given by the cosmic $\gamma$-ray horizon (CGRH), which has not been clearly observed yet. The CGRH is by definition the energy, $E_{0}$, at which the optical depth of the photon-photon pair production becomes unity as a function of redshift. Therefore, it gives an estimate of how far VHE photons can travel through the Universe. Due to the exponential behavior of the flux attenuation, an alternative definition is the energy at which the intrinsic spectrum is attenuated by the EBL by a factor of $1/e$ (see \eg \citealt{aharonian04}). (The intrinsic source spectrum is the one that we would observe if there were no effect from the EBL, also known as the EBL-corrected spectrum.)

An extreme category of AGNs, known as blazars, has been shown to be the best target for extragalactic VHE detections. They are characterized by having their energetic $\gamma$-ray jets pointing towards us. In fact, most of the extragalactic sources already detected by imaging atmospheric Cherenkov telescopes (IACTs) such as H.E.S.S., MAGIC, and VERITAS (\citealt{hinton04,lorenz04,weekes02}, respectively) are blazars\footnote{See for an updated compilation: http://tevcat.uchicago.edu/}. The observations of broadband spectral energy distributions (SEDs) of blazars show that they are characterized by a double-peaked shape and that their emission covers all the electromagnetic spectrum from radio up to the most energetic $\gamma$-rays. The synchrotron/synchrotron self-Compton (SSC) model provides a successful explanation for this behavior for most cases (\eg \citealt{abdo11b,abdo11d,zhang12}). In this framework a population of ultra-relativistic electrons causes the lower energy peak by synchrotron emission, while the second peak is then accounted for by inverse Compton production of $\gamma$-rays from the same population of high energy electrons and photons in the low energy peak.

The direct observation of the CGRH in the VHE spectra of blazars remains elusive. This is due to two main observational difficulties. First, the lack of knowledge of the intrinsic spectra at VHE. Extensive multiwavelength campaigns from radio up to $\gamma$-rays are needed in order to predict the unattenuated VHE emission from the synchrotron/SSC model with enough precision. These campaigns should preferably be simultaneous due to the short-time flux variability of blazars (\eg \citealt{aleksic11a,aleksic11b}). In this situation, the typical procedure in the literature to estimate the intrinsic VHE spectrum is either to assume a limit for the hardness of the slope $E^{-\Gamma}$ with $\Gamma=1.5$ (\citealt{aharonian06}) or to extrapolate the \emph{Fermi}-Large Area Telescope (LAT) spectrum up to higher energies (\citealt{georganopoulos10,orr11}).

Second, it has been possible only in recent years to detect a considerable number of blazars in the GeV energy range to allow us a statistical analysis. A large sample is necessary to reject intrinsic behaviors in the sources that mimic the effect of the CGRH. This improvement has been made thanks to the large data sets provided by the \emph{Fermi} satellite (\citealt{ackermann11}) and the IACTs.

The only estimations of the CGRH so far are EBL-model-dependent lower limits from VHE observations of blazars (\citealt{albert08}), lower limits from \emph{Fermi}-LAT observations of blazars (\citealt{abdo10b}), and the predictions from EBL models (\eg \citealt{franceschini08,kneiske10,finke10,dominguez11a}, hereafter D11; \citealt{gilmore12,stecker12}). (A table with a classification and description of the ingredients of these EBL models can be found in the proceeding by \citealt{dominguez12}.) Indeed, an independent observation of the CGRH will also provide a completely independent and new test to the modeling of the EBL and consequently constraints on galaxy evolution. Furthermore, the CGRH measurement also can be useful to estimate the cosmological parameters with a novel and independent methodology (\citealt{blanch05a,blanch05b,blanch05c}; Dom\'inguez \& Prada, submitted). The detection of the CGRH is a primary scientific goal of the \emph{Fermi} $\gamma$-ray Telescope (\citealt{hartmann07}).



This paper is organized as follows. Section~\ref{sec:data} describes the blazar catalog used in our analysis. The methodology is explained in Section~\ref{sec:methodology}. Section~\ref{sec:results} shows the results obtained from our analysis and in Section~\ref{sec:discussion} the results are discussed. Finally, a brief summary of the main results is presented in Section~\ref{sec:summary}.

Throughout this paper a standard $\Lambda$CDM cosmology is assumed, with $\Omega_{m}=0.3$, $\Omega_{\Lambda}=0.7$, and $H_{0}=70$~km~s$^{-1}$~Mpc$^{-1}$ (\citealt{larson11,komatsu11}).

\section{Data set}
\label{sec:data}
A catalog of quasi-simultaneous multiwavelength data from radio up to VHE for 15 blazars has been built. The data for energies lower than the \emph{Fermi}-LAT regime (20~MeV--$>$300~GeV; \citealt{atwood09}) are taken from the data compilation presented by \citet{zhang12}. We refer the reader to that paper for references and details of the different data sets and their simultaneity. That catalog is combined here with \emph{Fermi}-LAT data from the second \emph{Fermi}-LAT AGN catalog (2FGL, \citealt{ackermann11}) and the new \emph{Fermi}-LAT hard-spectrum catalog (1FHL, Ackermann et al., in preparation), which include two and three years of \emph{Fermi}-LAT data, respectively. The second \emph{Fermi} AGN catalog contains fluxes in the following six energy bins: 30--100~MeV, 100--300~MeV, 300~MeV--1~GeV, 1--3~GeV, 3--10~GeV and 10--100~GeV. The \emph{Fermi} hard-spectrum catalog contains fluxes for the sources with the highest-energy emission in three bins that reach higher energy than the ones used in the second-year catalog (10--30~GeV, 30--100~GeV and 100--500~GeV). Our final catalog includes data from IACTs as well. We use simultaneous VHE data from IACTs when available; otherwise the IACT observation closest in time to the lower-energy data is used following the suggestions by \citet{zhang12}. Table~\ref{tab1} lists the 15 sources in the set of blazars that we study here. The catalog presented in \citet{zhang12} contains 24 blazars, however we could not use all of them due to non-detections either by \emph{Fermi} or the IACTs, which are essential for applying our methodology (see Section~\ref{sec:methodology}).

Our catalog covers a wide redshift range from $z=0.031$ to $z\sim 0.5$ and all 15 blazars are classified as BL-Lac sources (which are typically characterized by rapid and large-amplitude flux variability and significant optical polarization). We note that in the cases of PG~1553+113 and 3C~66A the redshifts are uncertain, but still these sources are included in the analysis. A redshift in the range $0.395<z\le 0.58$ is estimated for PG~1553+113 by Danforth et al. (2010, the upper limit is $1\sigma$). The blazar 3C 66A typically is cited as having a redshift of $z=0.444$, which is used here as well (cf. \citealt{bramel05,finke08a}).

\section{Methodology}
\label{sec:methodology}
Our methodology consists of finding the best-fitting homogeneous synchrotron/SSC models from multiwavelength data as simultaneous as possible from radio to the highest-energy $\gamma$-rays detected by the \emph{Fermi}-LAT for the blazars in our catalog. These models predict the unattenuated VHE fluxes, which are compared with detections by IACTs. The ratios between the unattenuated and detected VHE fluxes give an estimate of the EBL optical depth. By means of a maximum likelihood technique that is independent of any EBL model and that is based only on a few physically motivated assumptions, we then derive the CGRH for each blazar.


\subsection{Broadband spectral-energy-distribution fitting and optical-depth data estimation}
\begin{figure*}
\includegraphics[trim=1.7cm 0 2cm 0.8cm,clip=True,width=\columnwidth]{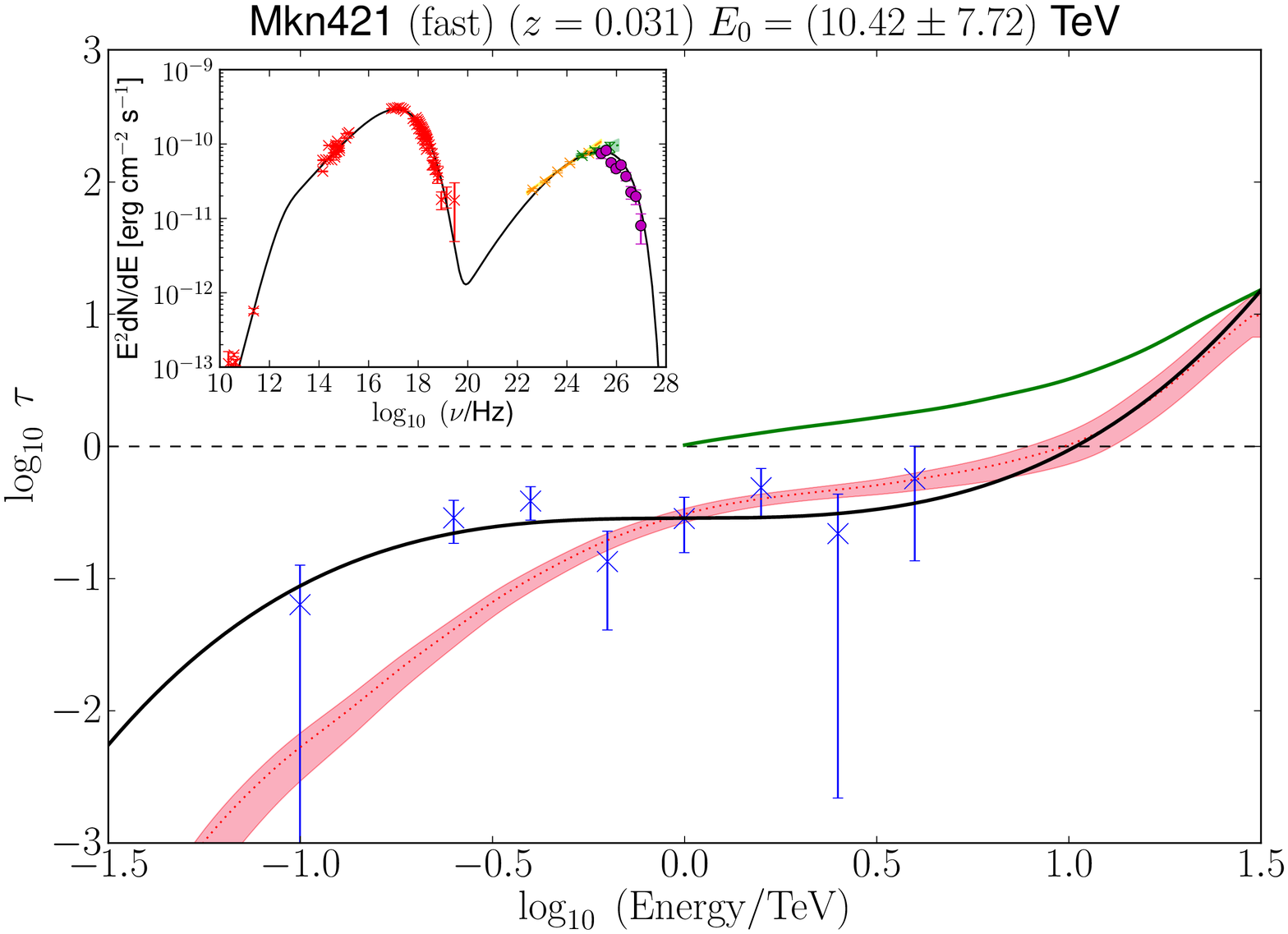}
\includegraphics[trim=1.7cm 0 2cm 0.8cm,clip=True,width=\columnwidth]{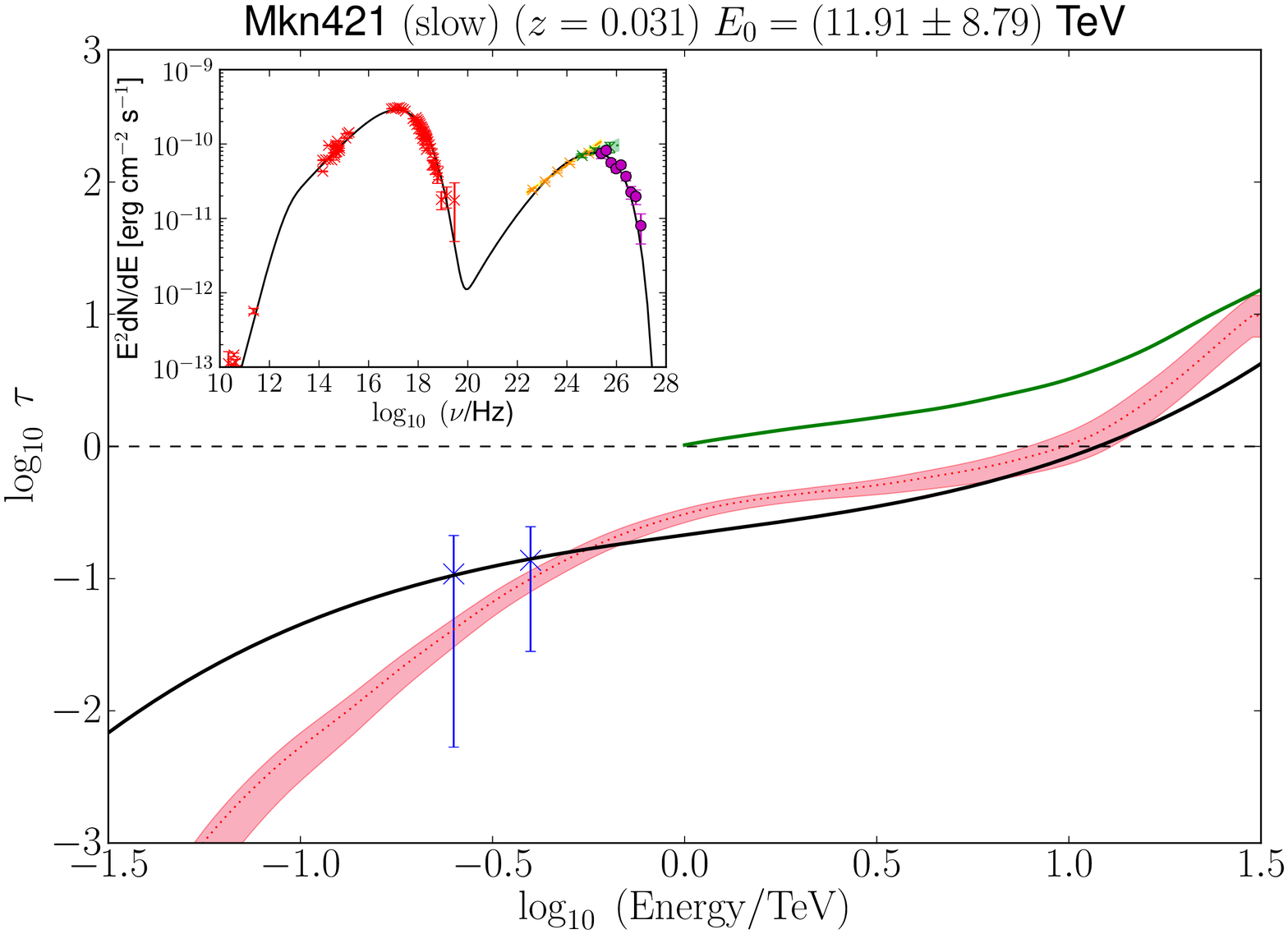}\\
\includegraphics[trim=1.7cm 0 2cm 0.8cm,clip=True,width=\columnwidth]{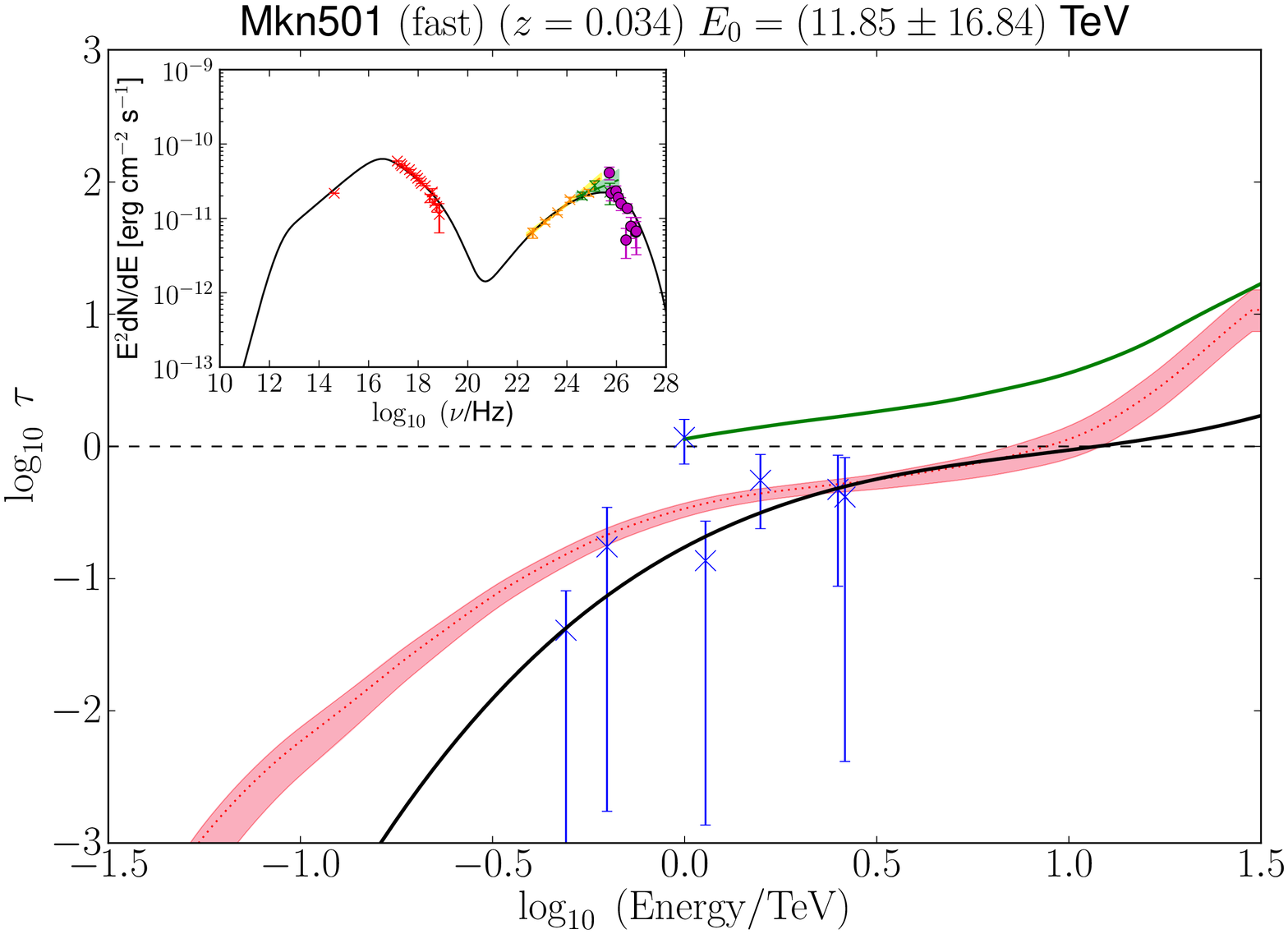}
\includegraphics[trim=1.7cm 0 2cm 0.8cm,clip=True,width=\columnwidth]{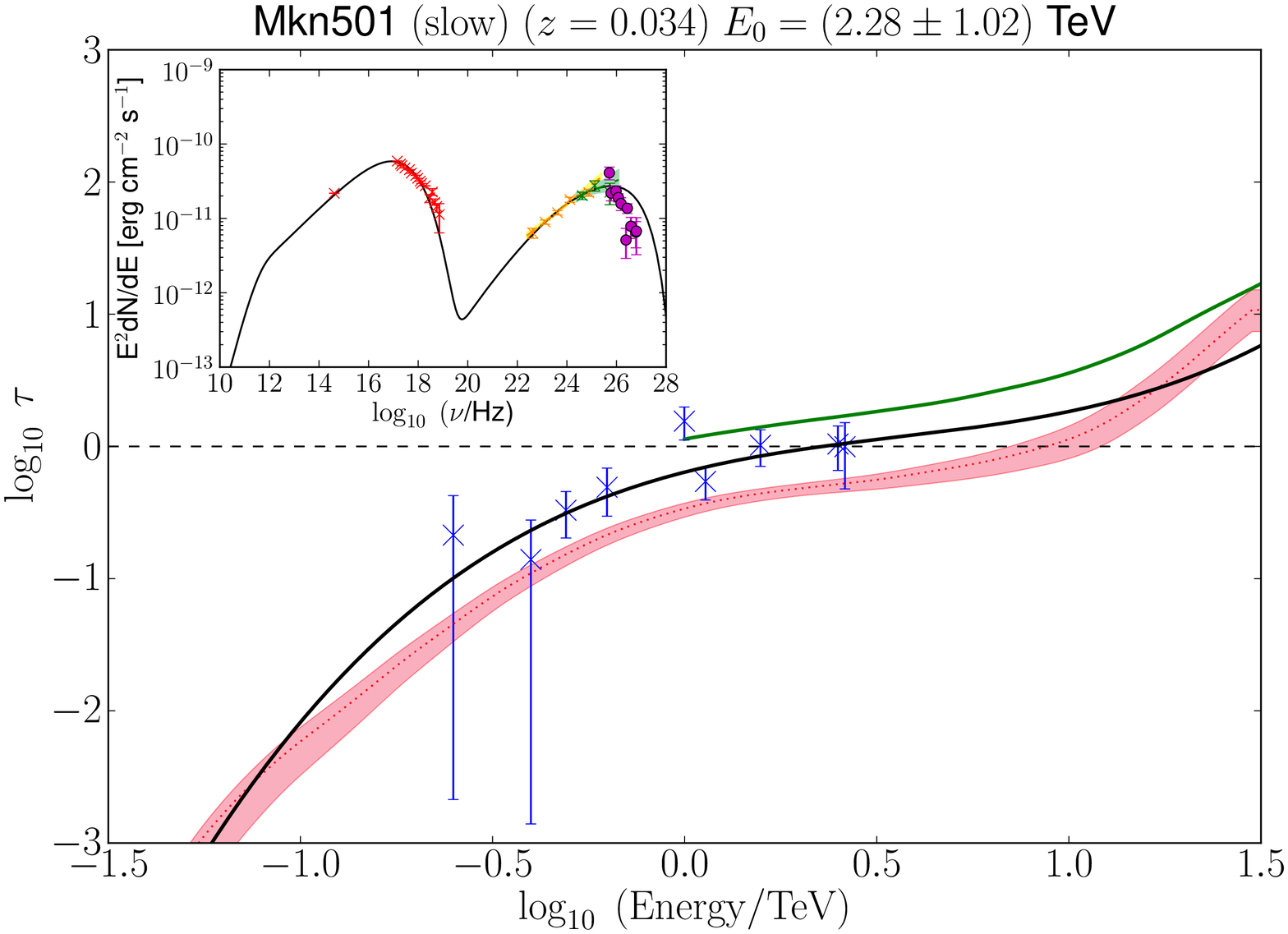}\\
\includegraphics[trim=1.7cm 0 2cm 0.8cm,clip=True,width=\columnwidth]{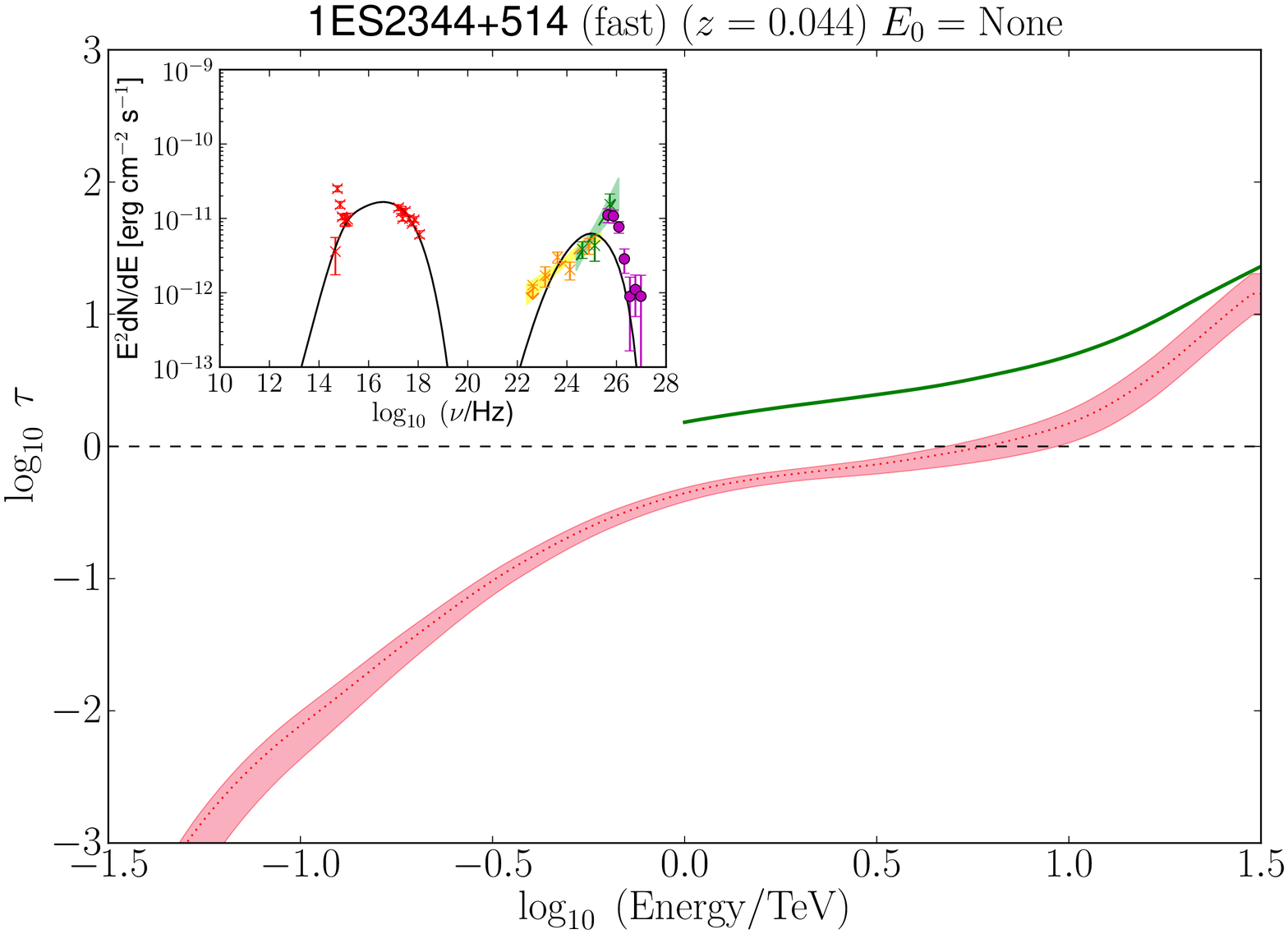}
\includegraphics[trim=1.7cm 0 2cm 0.8cm,clip=True,width=\columnwidth]{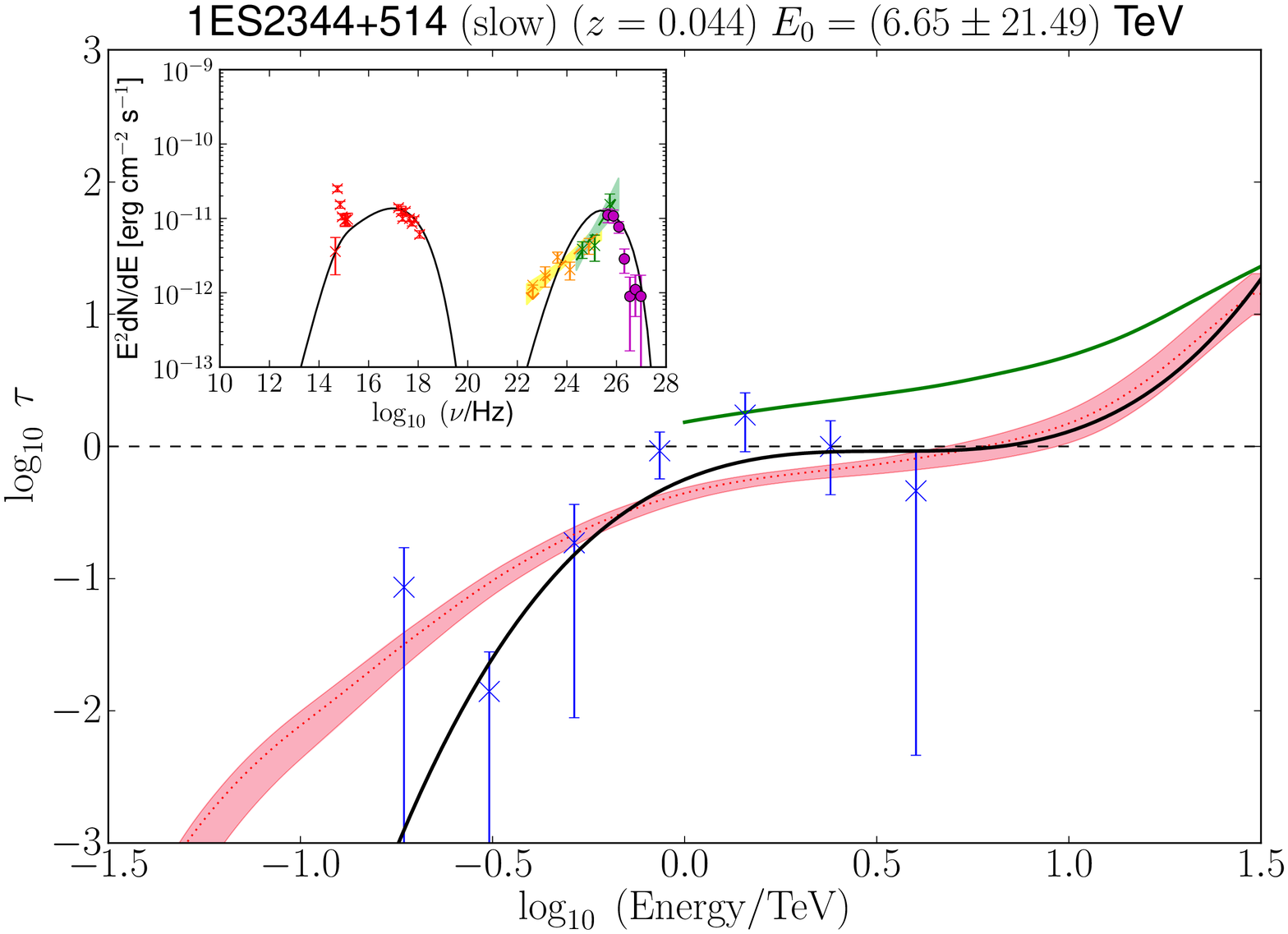}\\
\caption{The optical-depth data estimated from equation~\ref{eq:tau} for our sample of blazars (slow and fast variability timescales) are shown with blue crosses in order of increasing redshift. The most likely polynomial is shown with a solid-black line. If there are no polynomials in the figure it is because the observed VHE fluxes are higher than the prediction by the synchrotron/SSC model (probably due to simultaneity issues; see section~\ref{sec:discussion}), which leads to no optical depth data. It may happen that none of the polynomials satisfied our boundary conditions as well. In those cases no solution for the CGRH is found. The optical-depth estimation over redshift from the EBL model discussed in \citet{dominguez11a} is shown for comparison with a red band that include its uncertainties. The solid-green line shows the upper limits of the optical depth derived from the EBL upper limits found in \citet{mazin07}. The $\log_{10}(\tau)=0$ is marked as a dashed line in Figure~\ref{fig:E0fits} to guide the reader's eye. The synchrotron-self Compton fit for each blazar is shown in each panel as a inset figure with the multiwavelength data (upper limits are shown with arrows pointing downwards). The lower energy data is shown with red crosses (\citealt{zhang12}), the \emph{Fermi}-LAT data from the second-year public catalog are shown with orange color (\citealt{ackermann11}) and its uncertainties with a \emph{butterfly}, the LAT data from the hard-source catalog are shown with green color (Ackermann et al., in preparation) and the IACT data are shown in magenta (see references in Table~\ref{tab2}). The left column shows the results for the fast minimum time variability SSC model ($t_{v,min}=10^{4}$~s) whereas the right column shows the results for the slow model ($t_{v,min}=10^{5}$~s). The name of the blazar, the minimum time variability, its redshift and the CGRH ($E_{0}$) derived from every fit are listed in the title of each panel.}
\label{fig:E0fits}
\end{figure*}

For every blazar in our sample, we built a quasi-simultaneous SED based on the data collected by \citet{zhang12} and on the LAT data from the second \emph{Fermi}-LAT AGN catalog (\citealt{ackermann11}) and the hard source catalog (Ackermann et al., in preparation). These data are shown in the insets of Figure~\ref{fig:E0fits}. In many cases, these SEDs are not constructed from strictly simultaneous data. However, since we preferentially choose SEDs that are for a low state, we expect that the effects of variability are minimal in the $\gamma$-ray energy range. We then fit the unique multiwavelength data of each source with a synchrotron/SSC model using a $\chi^{2}$ minimization technique. The model and fitting technique are fully described in Finke, Dermer \& B\"ottcher (2008, see also \citealt{mankuzhiyil10}). This model includes photoabsorption by photons internal to the blob.  Since BL Lac objects and their presumed misaligned counterparts, FR~I radio galaxies usually lack optically thick dust tori (\citealt{donato04,plotkin12}) we do not include photoabsorption from this radiation source, although it could in principle be important. We note that for 1ES~1959+650, there is evidence for an optically thick dust component (\citealt{falomo00}). However, we did not obtain positive results with this source, as discussed in Section~\ref{sec:discussion} below. The fitting technique is double nested, with the \emph{inner loop} fitting the synchrotron component with a particular electron distribution. In this paper, we use a broken power-law for the electron distribution with an exponential cutoff at high energies, that is, at $\gamma_{max}$. In some cases (Mrk~421, 1ES~2344+514, PKS~2005$-$489, H~2356$-$309, 1ES~218+304, and 1ES~1101$-$232) we found that a single power-law with exponential cutoff was sufficient to provide good fits. Our fits have as free parameters the electron indices $p_1$, $p_2$; the minimum, maximum, and break electron Lorentz factors, $\gamma_{min}$, $\gamma_{max}$, and $\gamma_{brk}$, respectively; and the overall normalization.  Often $\gamma_{min}$ and $\gamma_{max}$ were kept constant during the fit. The \emph{outer loop} fit the SSC model the high energy data, and has three free parameters: the Doppler factor, $\delta_{D}$; the magnetic field strength, $B$; and the minimum variability timescale, $t_{v,min}$. We assume that the bulk Lorentz factor $\Gamma_{bulk} = \delta_{D}$. In the fits, $t_{v,min}$ was kept constant, with only $\delta_{D}$ and $B$ as free parameters. We discuss in more detail our choices of $t_{v,min}$ below. In the fits, we specifically leave off the IACT data; we only fit the IR through the LAT $\gamma$-rays. We use any radio points as upper limits, since this emission is likely from another region of the jet. We leave out the IACT data because for this fit, we are fitting data which are unaffected by EBL attenuation. In some of the more distant sources, the highest energy LAT point (at energy $\sim 224$~GeV) can suffer significant attenuation. Therefore, we remove this data point for sources at $z>0.05$. This choice is supported by a variety of observational evidences (see \citealt{primack11,dominguez12}). Once we have the resulting model curve from our fit that includes the extrapolation to VHE energies (the overall model is shown with a black line in the insets of Figure~\ref{fig:E0fits}), we use this as the unattenuated/intrinsic spectrum for the source. We compare it with the flux observed from the IACT detection to calculate the absorption optical depth,

\begin{equation}
\label{eq:tau}
\tau(E,z)=\ln \Big (\frac{dF}{dE}\Big{|}_{int}/\frac{dF}{dE}\Big{|}_{obs}\Big )
\end{equation}
\noindent for photons observed in an energy bin centered at energy $E$ and a source at redshift $z$. Here $F_{obs}$ is the observed differential flux and $F_{int}$ is the intrinsic flux at the energies given by the IACT detection \ie the fluxes given by the synchrotron/SSC model evaluated at the energies sampled by the IACT. The uncertainties in $\tau$ come directly from the uncertainties in the IACT observations. The $\log_{10}(\tau)$ data are shown with blue crosses in Figure~\ref{fig:E0fits}. The method used here for measuring $\tau(E,z)$ is similar to the one described by \citet{mankuzhiyil10}.

For the synchrotron/SSC model, the radius of the spherical emitting region, $R_{blob}$ is determined from the minimum variability timescale, $t_{v,min}$, which is in turn constrained by the observed variability timescale through light travel time arguments so that $t_{v,min}\le t_{v}$  (\eg \citealt{finke08b}). For consistency, we used the same $t_{v,min}$ for all of our blazars, $t_{v,min}=10^{4}$~s and $t_{v,min}=10^{5}$~s.  As we see in Section~\ref{sec:results}, the choice of variability time makes very little difference to the model curve, although it has a large effect on the model fit parameters, which are not the focus of this paper. Thus we are confident that the choice of $t_{v,min}$ has little effect on our resulting measurement of $\tau(E,z)$. However, although the two SSC models are similar they predict different VHE fluxes, which allows us to include the uncertainty in the variability timescale in our analysis.

It should be noted that several sources have been observed to have extremely rapid ($\sim 10^2$~s) variability timescales (\eg \citealt{aharonian07}). These rapid flares are quite rare, and since we have chosen SEDs from blazars in a quiescent state, we think these short timescales (and the small emitting regions they imply) are unlikely. Furthermore, we have fit several of our objects with $t_{v,min}=10^2$~s and found the resulting Doppler factors to be extremely high, $\delta \ga$ a few hundred, and in two cases (1ES~1101$-$232 and 3C~66A) as high as 1000. We believe these high Doppler factors to be unreasonably high.

\begin{figure*}
\includegraphics[trim=1.7cm 0 2cm 0.8cm,clip=True,width=\columnwidth]{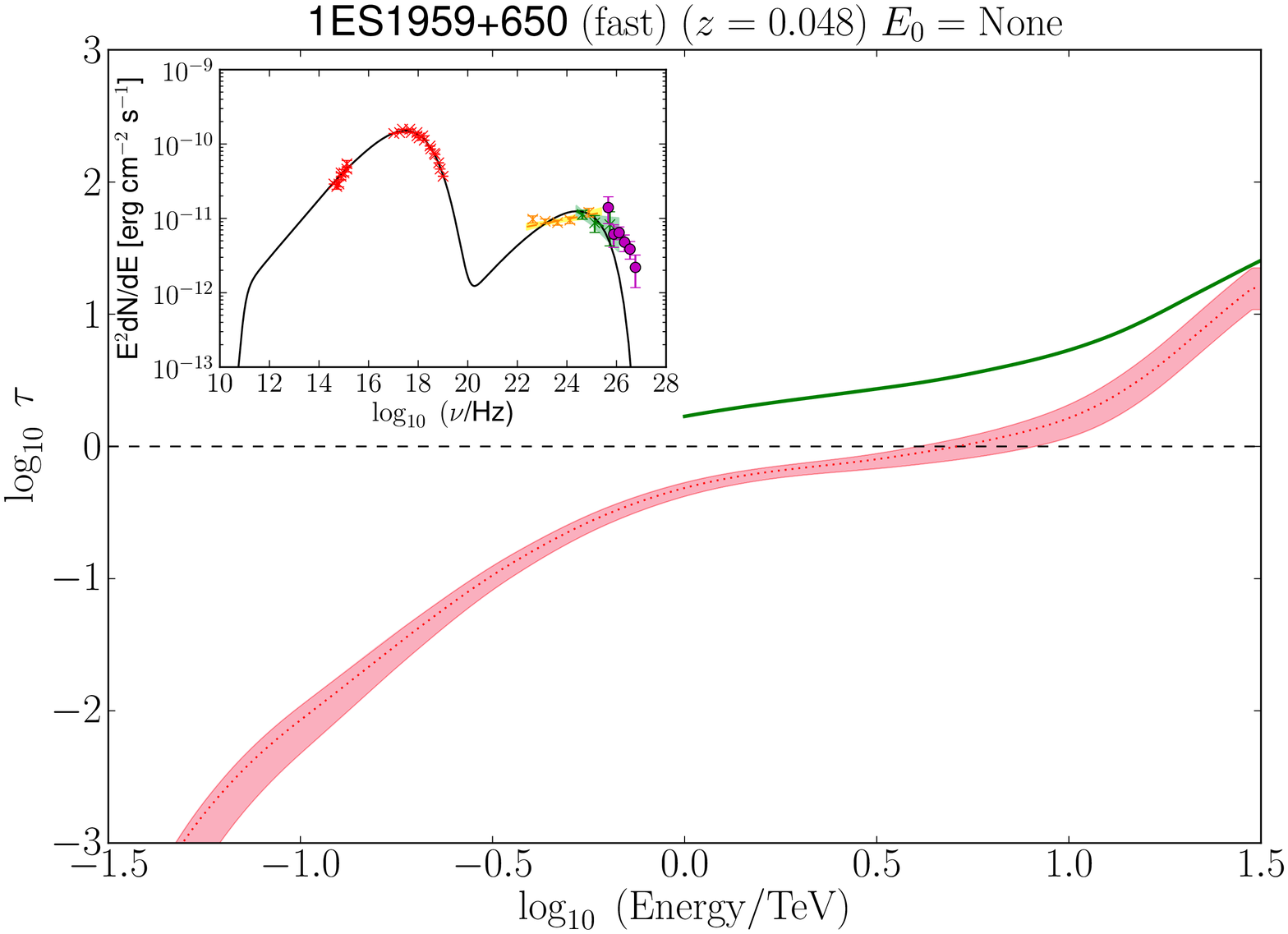}
\includegraphics[trim=1.7cm 0 2cm 0.8cm,clip=True,width=\columnwidth]{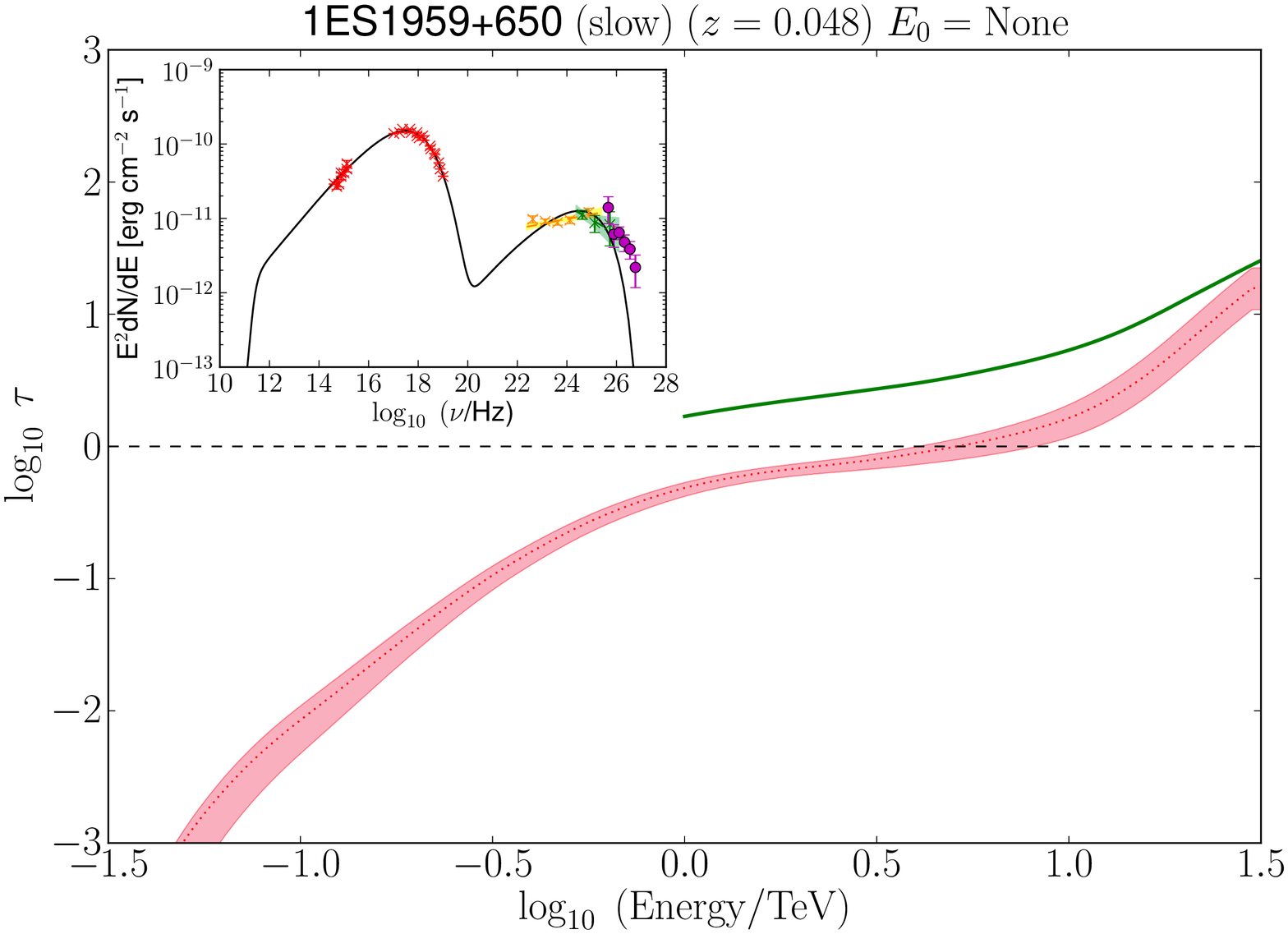}\\
\includegraphics[trim=1.7cm 0 2cm 0.8cm,clip=True,width=\columnwidth]{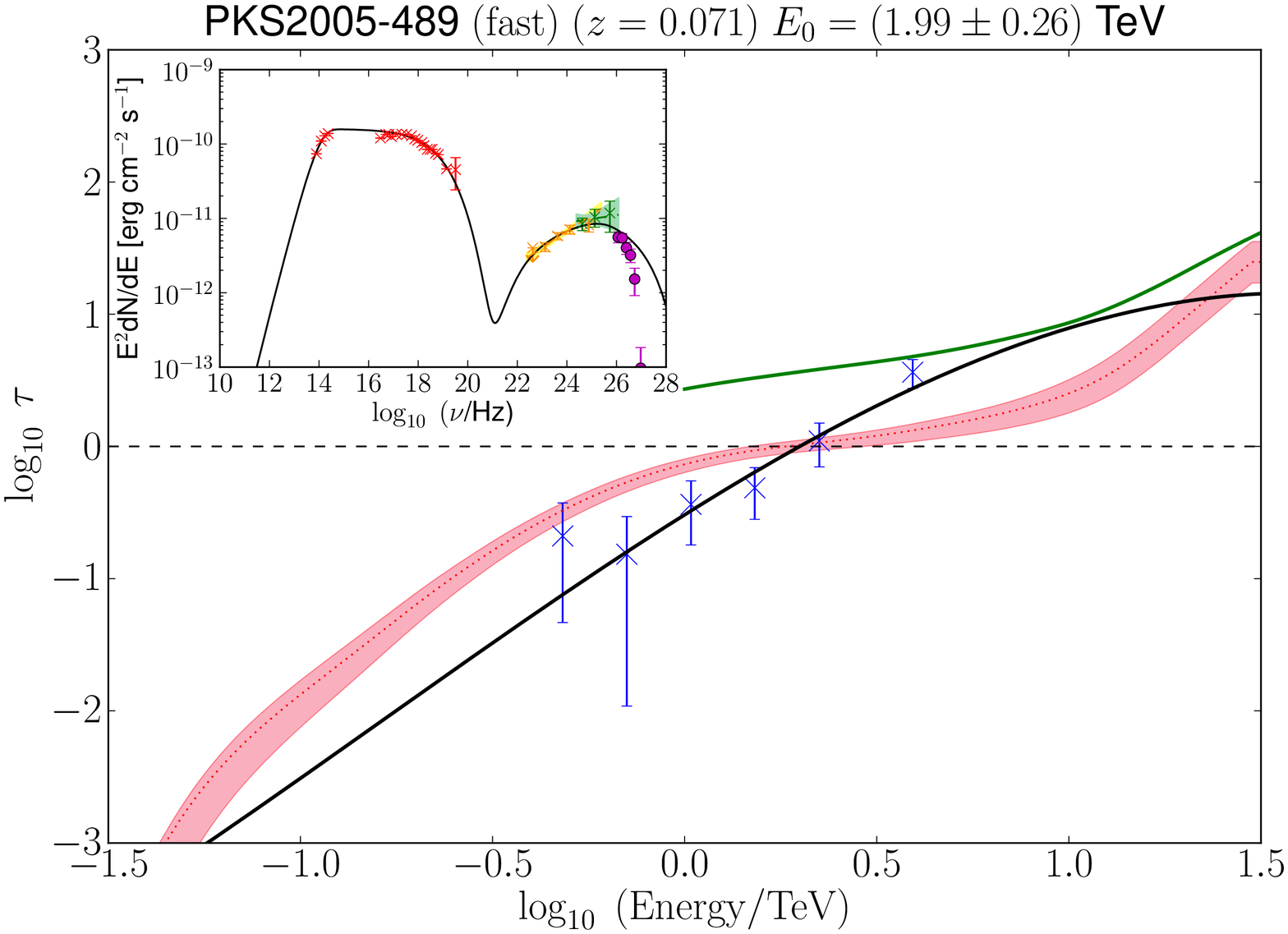}
\includegraphics[trim=1.7cm 0 2cm 0.8cm,clip=True,width=\columnwidth]{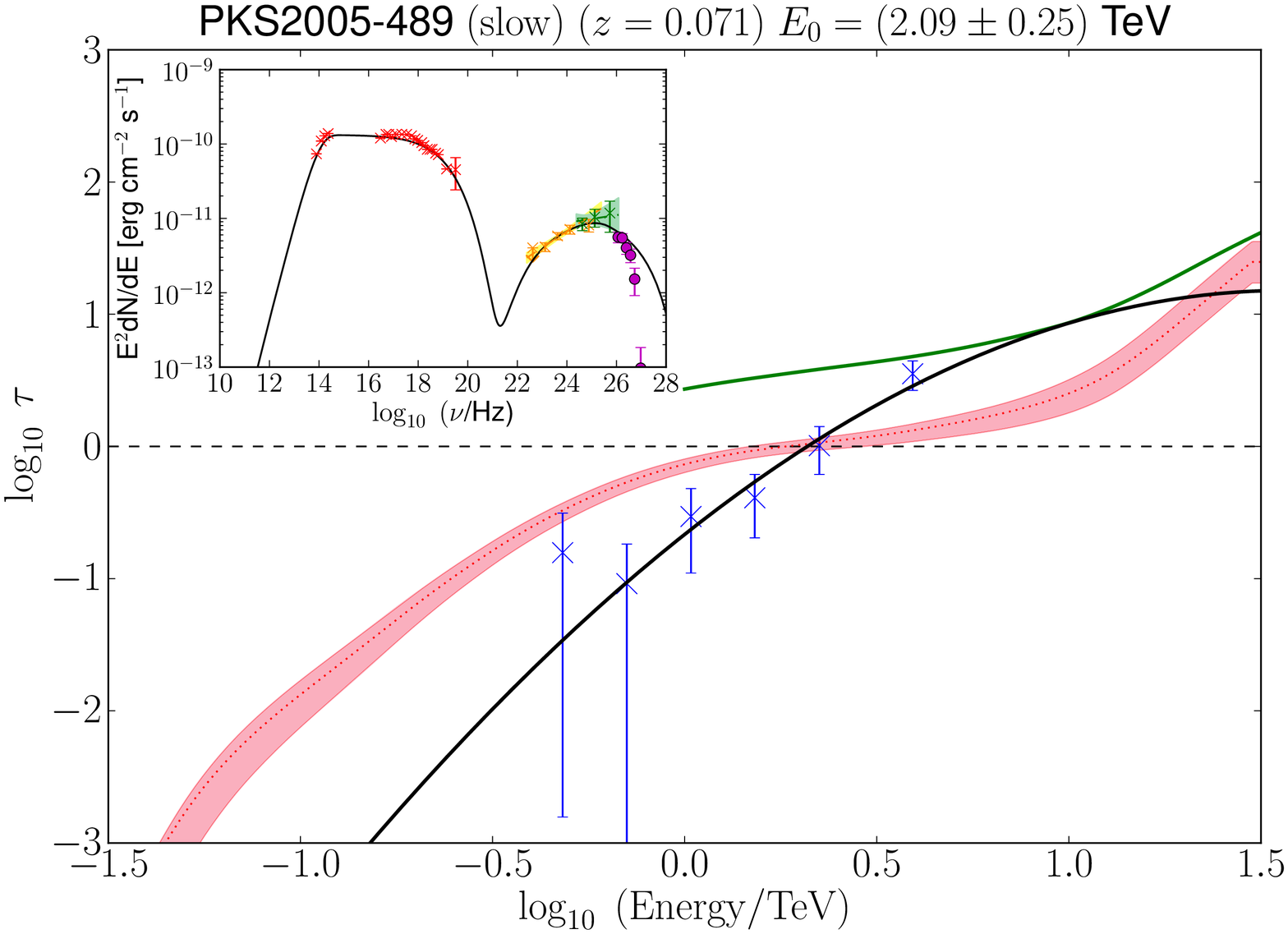}\\
\includegraphics[trim=1.7cm 0 2cm 0.8cm,clip=True,width=\columnwidth]{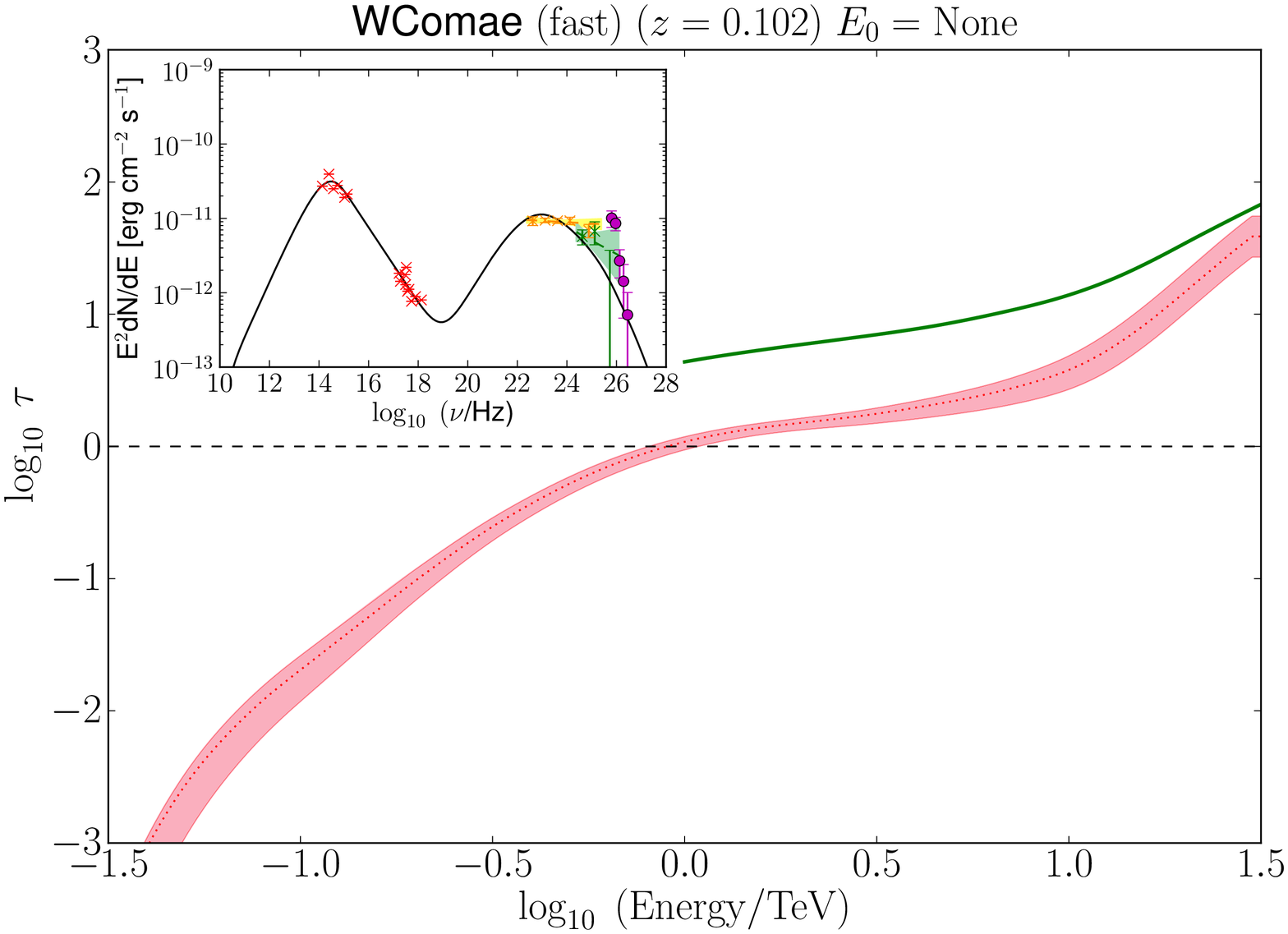}
\includegraphics[trim=1.7cm 0 2cm 0.8cm,clip=True,width=\columnwidth]{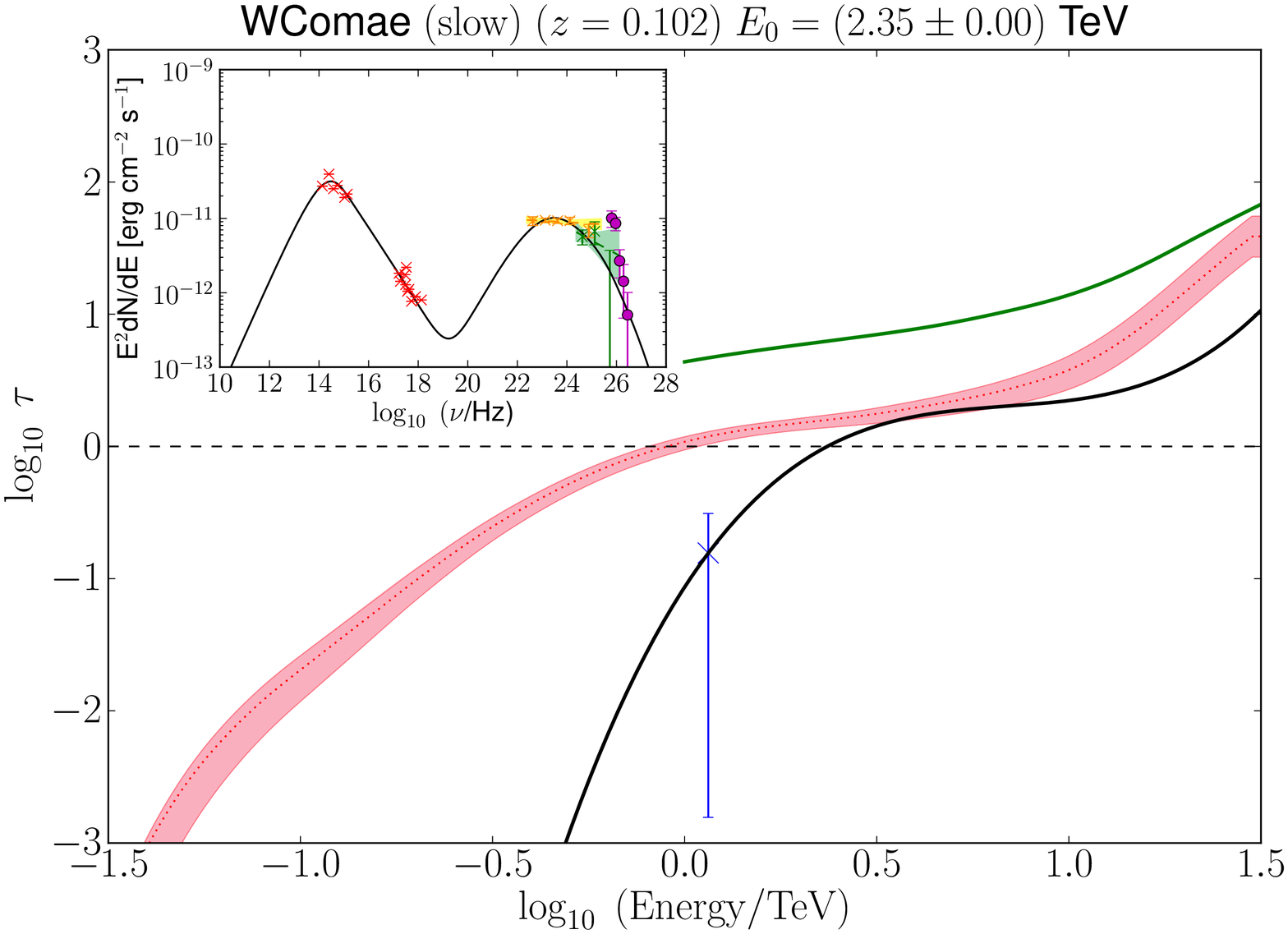}\\
\center{{\bf Figure 1.}---  continued}
\end{figure*}

\begin{figure*}
\includegraphics[trim=1.7cm 0 2cm 0.8cm,clip=True,width=\columnwidth]{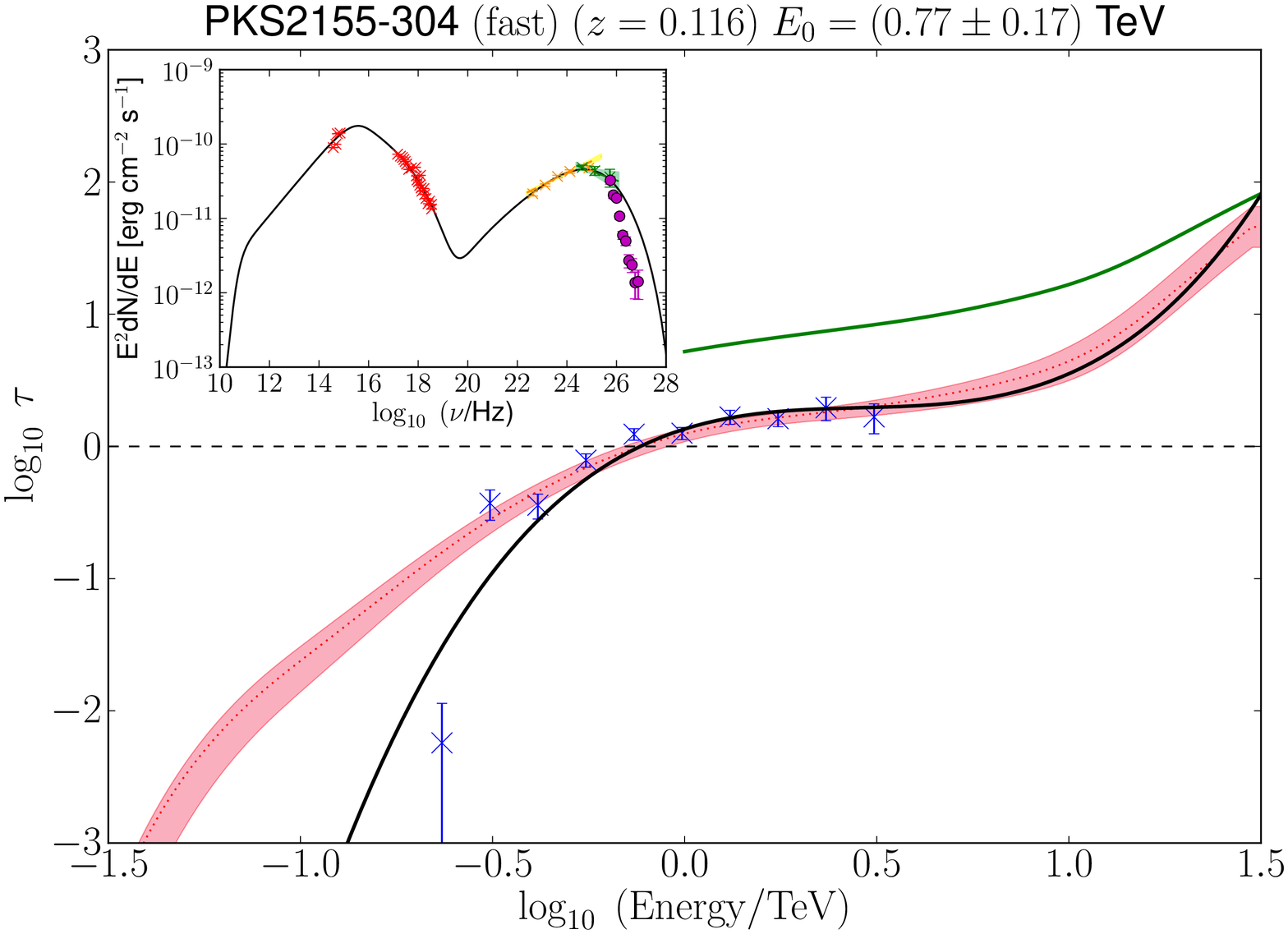}
\includegraphics[trim=1.7cm 0 2cm 0.8cm,clip=True,width=\columnwidth]{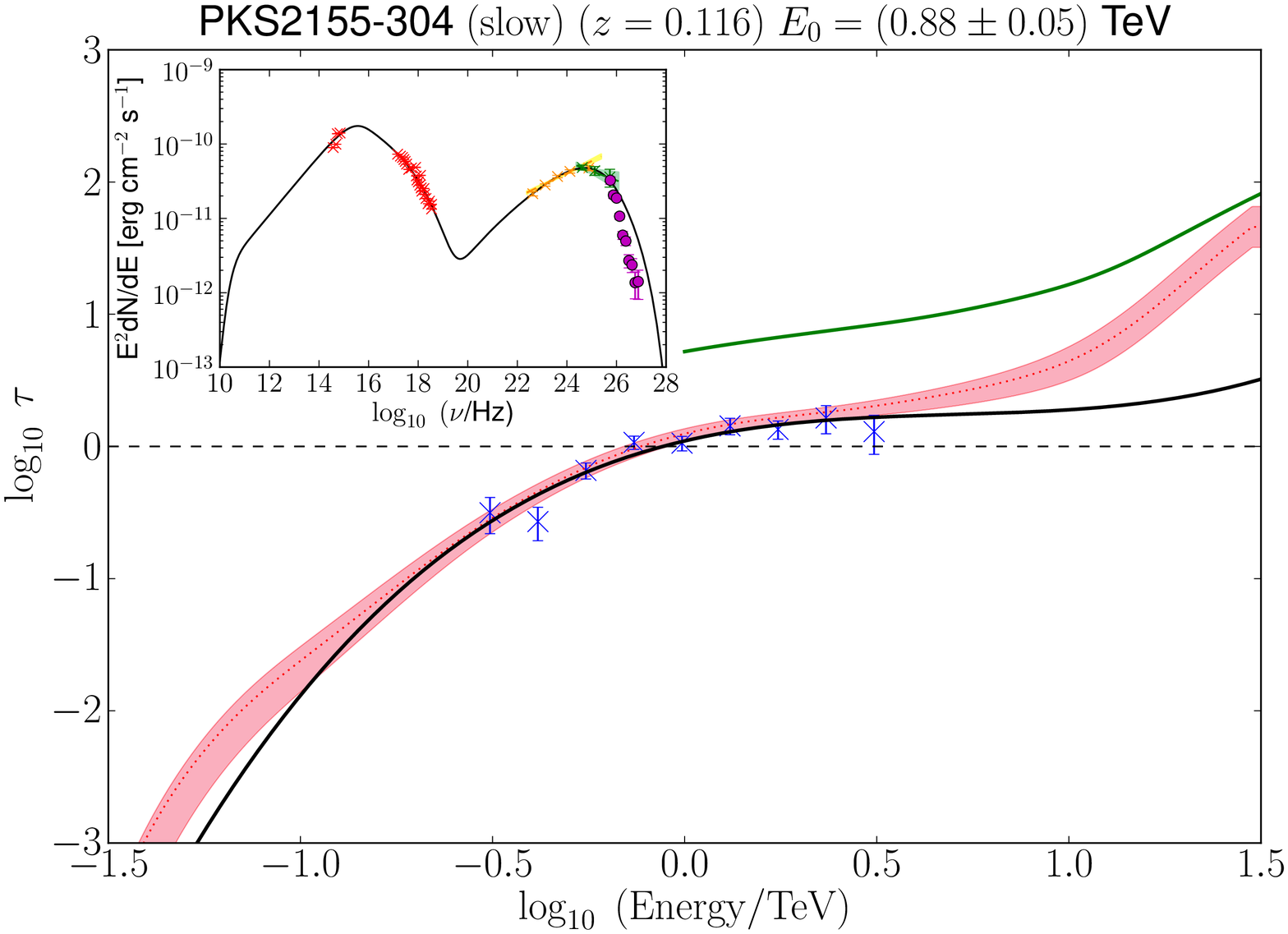}\\
\includegraphics[trim=1.7cm 0 2cm 0.8cm,clip=True,width=\columnwidth]{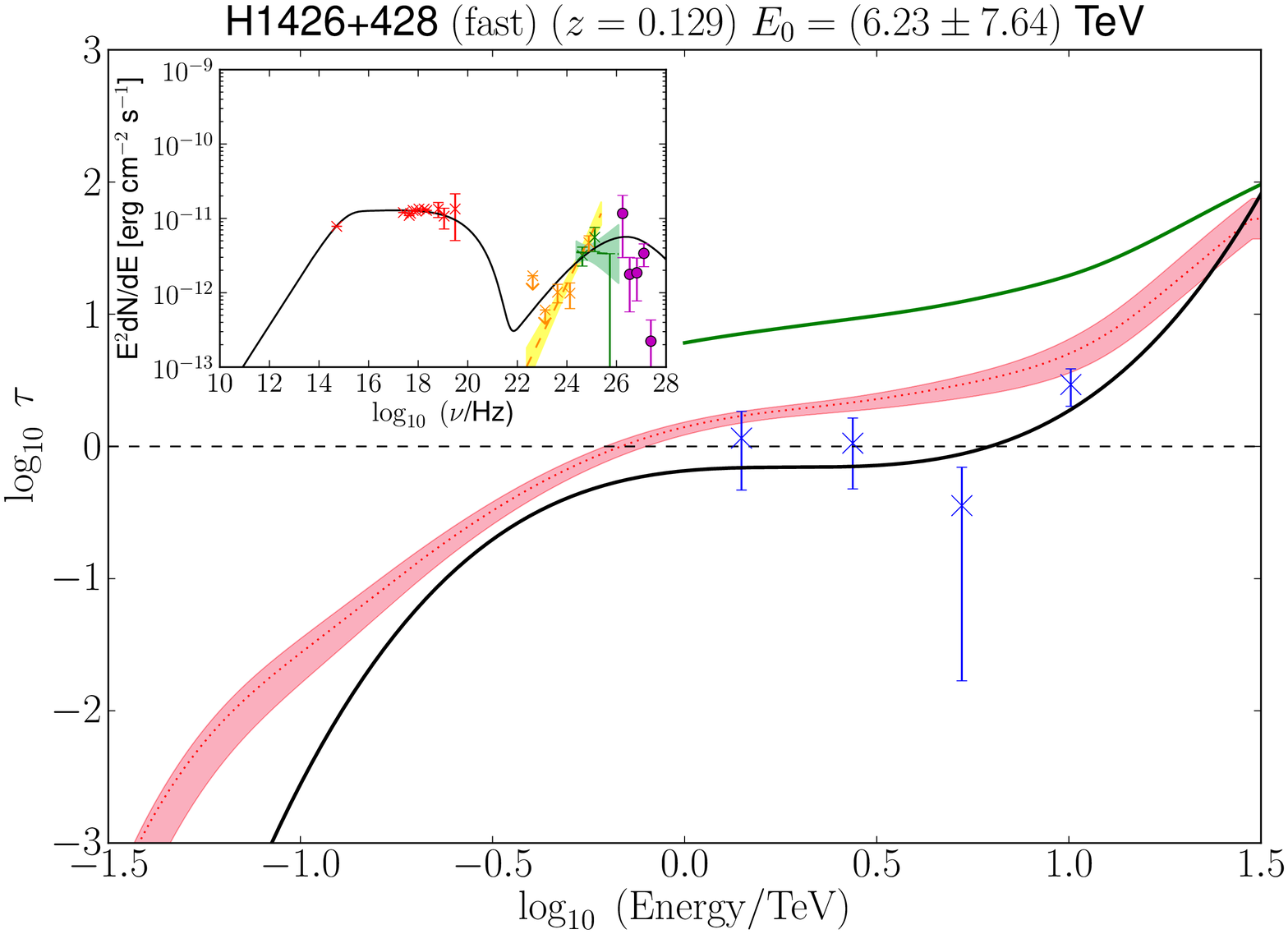}
\includegraphics[trim=1.7cm 0 2cm 0.8cm,clip=True,width=\columnwidth]{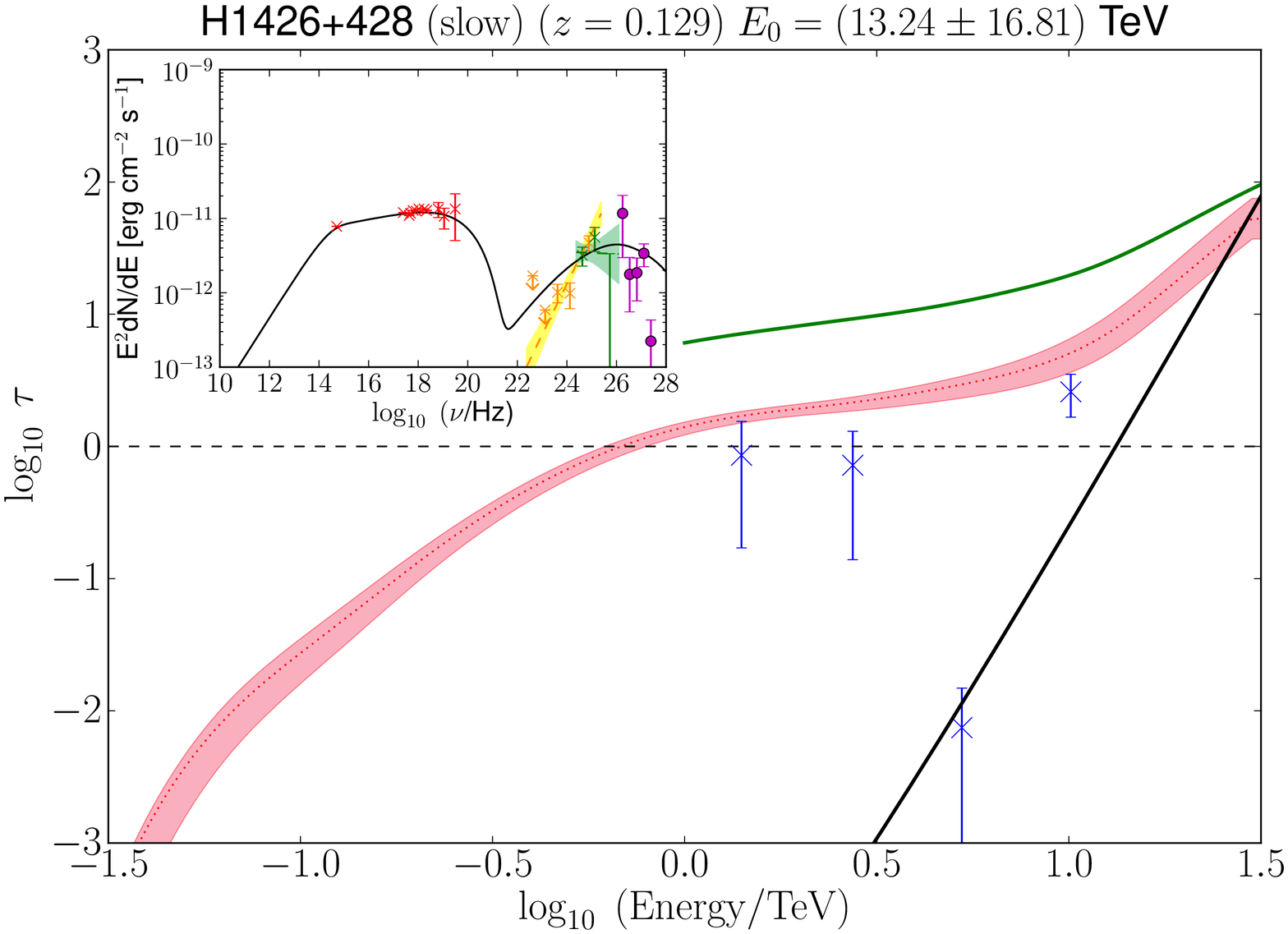}\\
\includegraphics[trim=1.7cm 0 2cm 0.8cm,clip=True,width=\columnwidth]{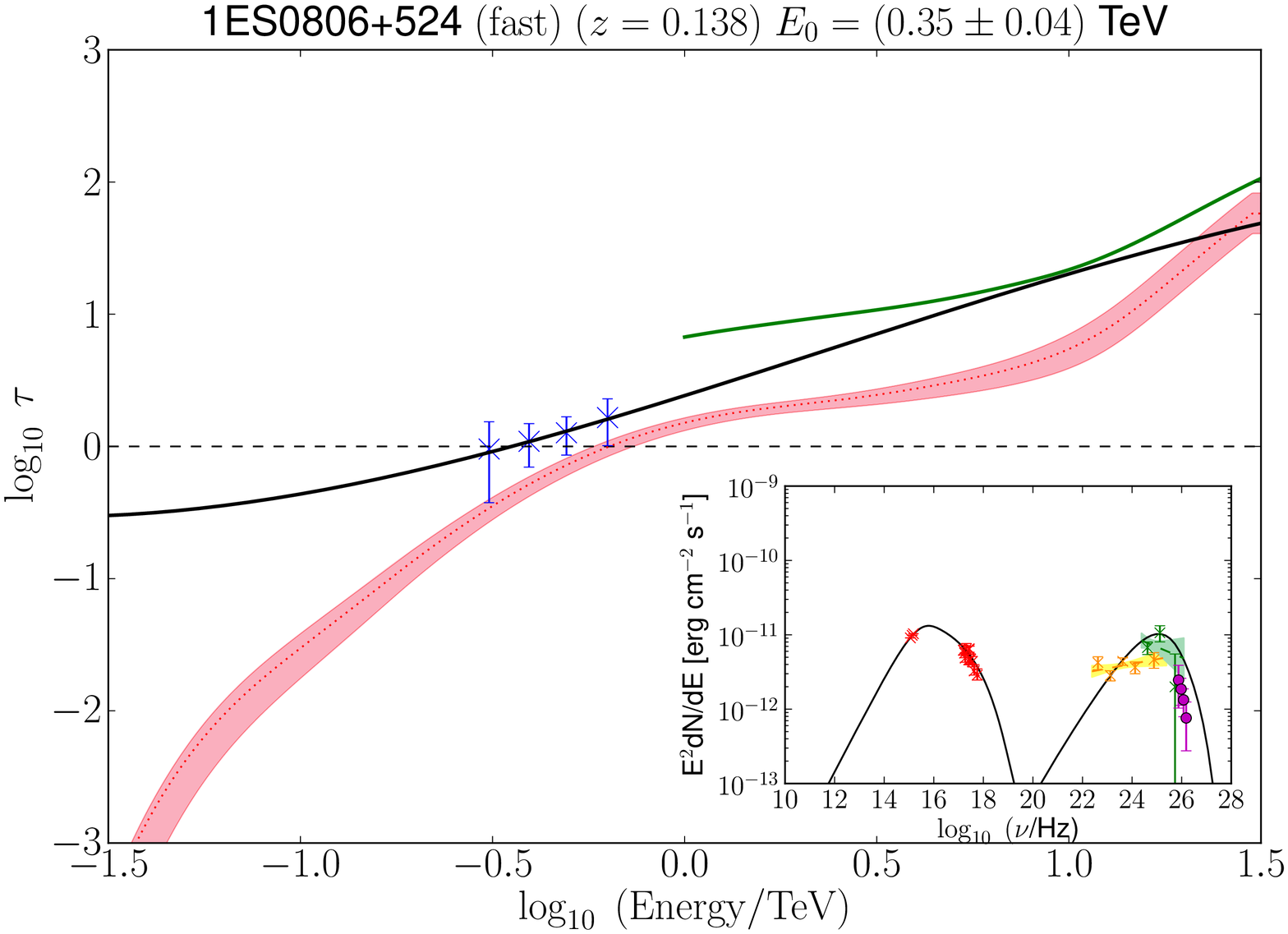}
\includegraphics[trim=1.7cm 0 2cm 0.8cm,clip=True,width=\columnwidth]{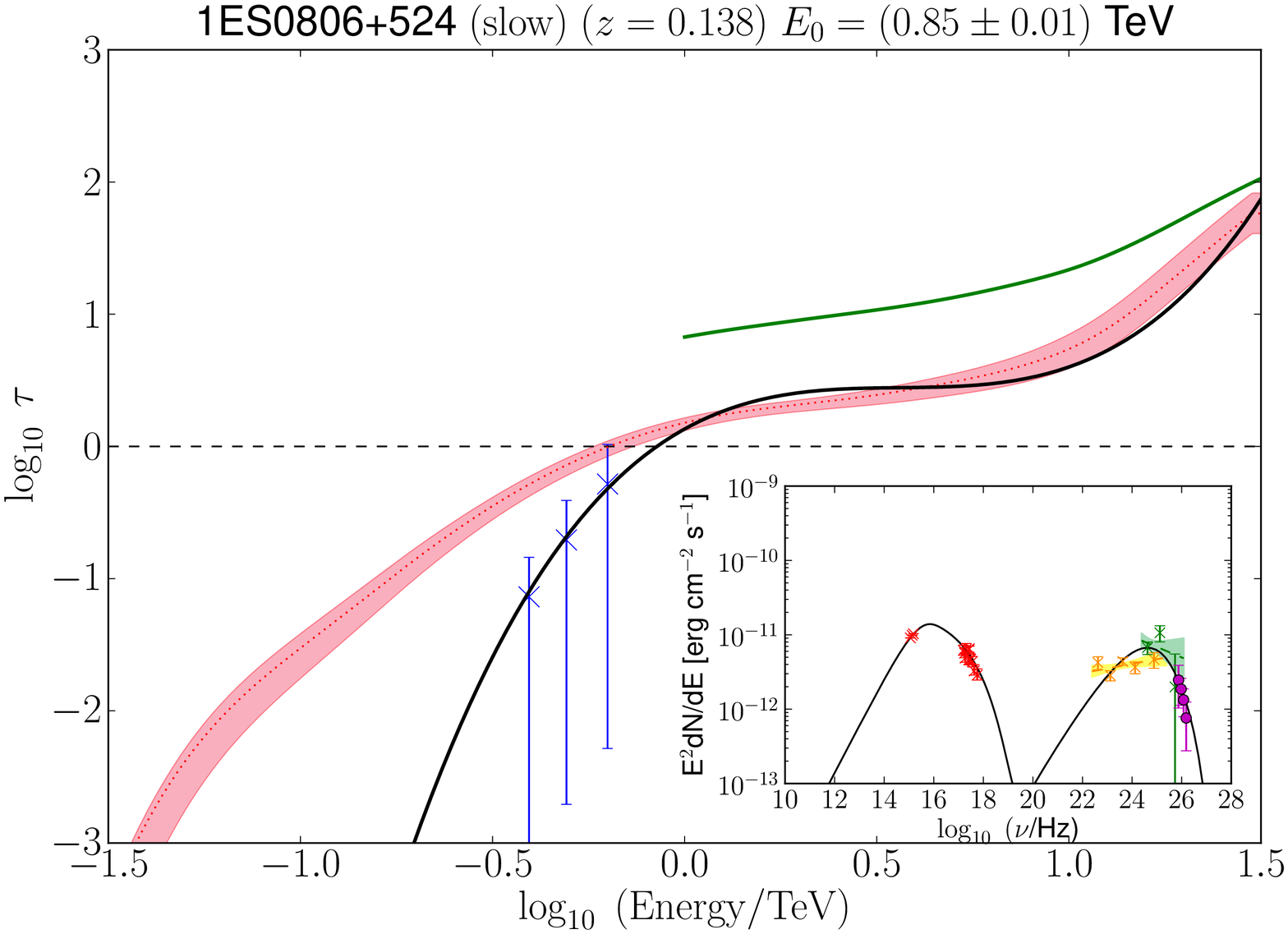}\\
\center{{\bf Figure 1.}---  continued}
\end{figure*}

\begin{figure*}
\includegraphics[trim=1.7cm 0 2cm 0.8cm,clip=True,width=\columnwidth]{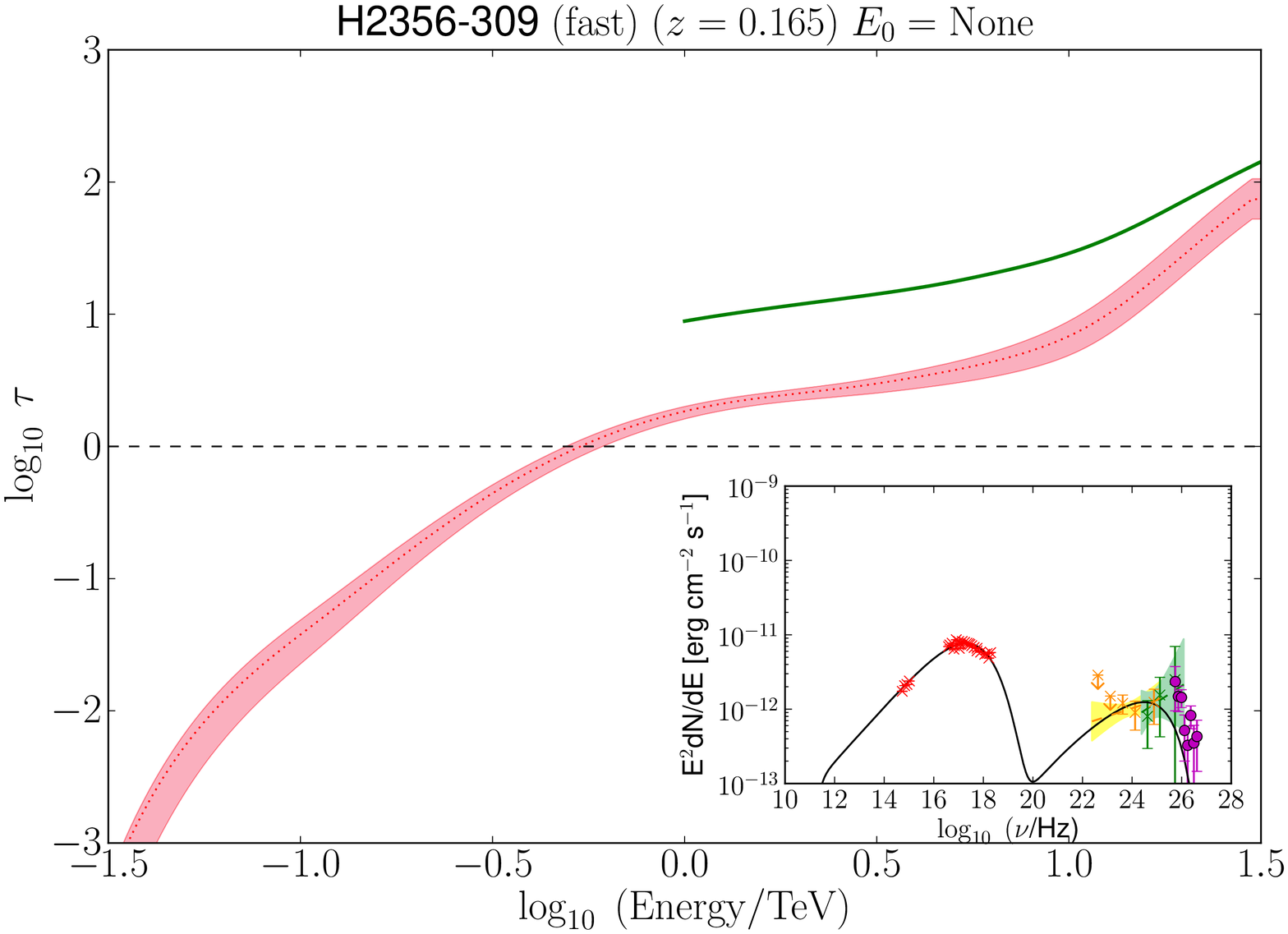}
\includegraphics[trim=1.7cm 0 2cm 0.8cm,clip=True,width=\columnwidth]{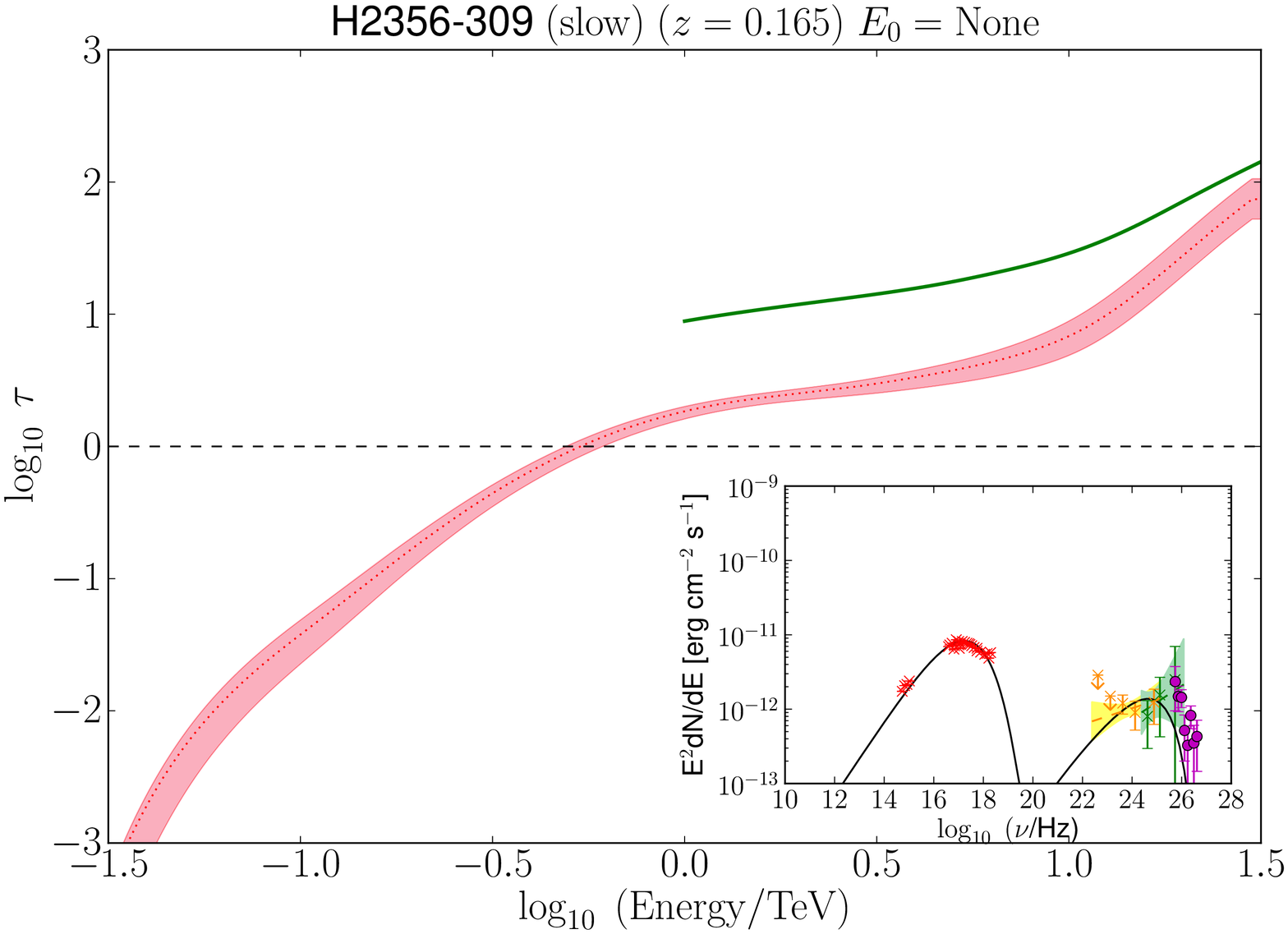}\\
\includegraphics[trim=1.7cm 0 2cm 0.8cm,clip=True,width=\columnwidth]{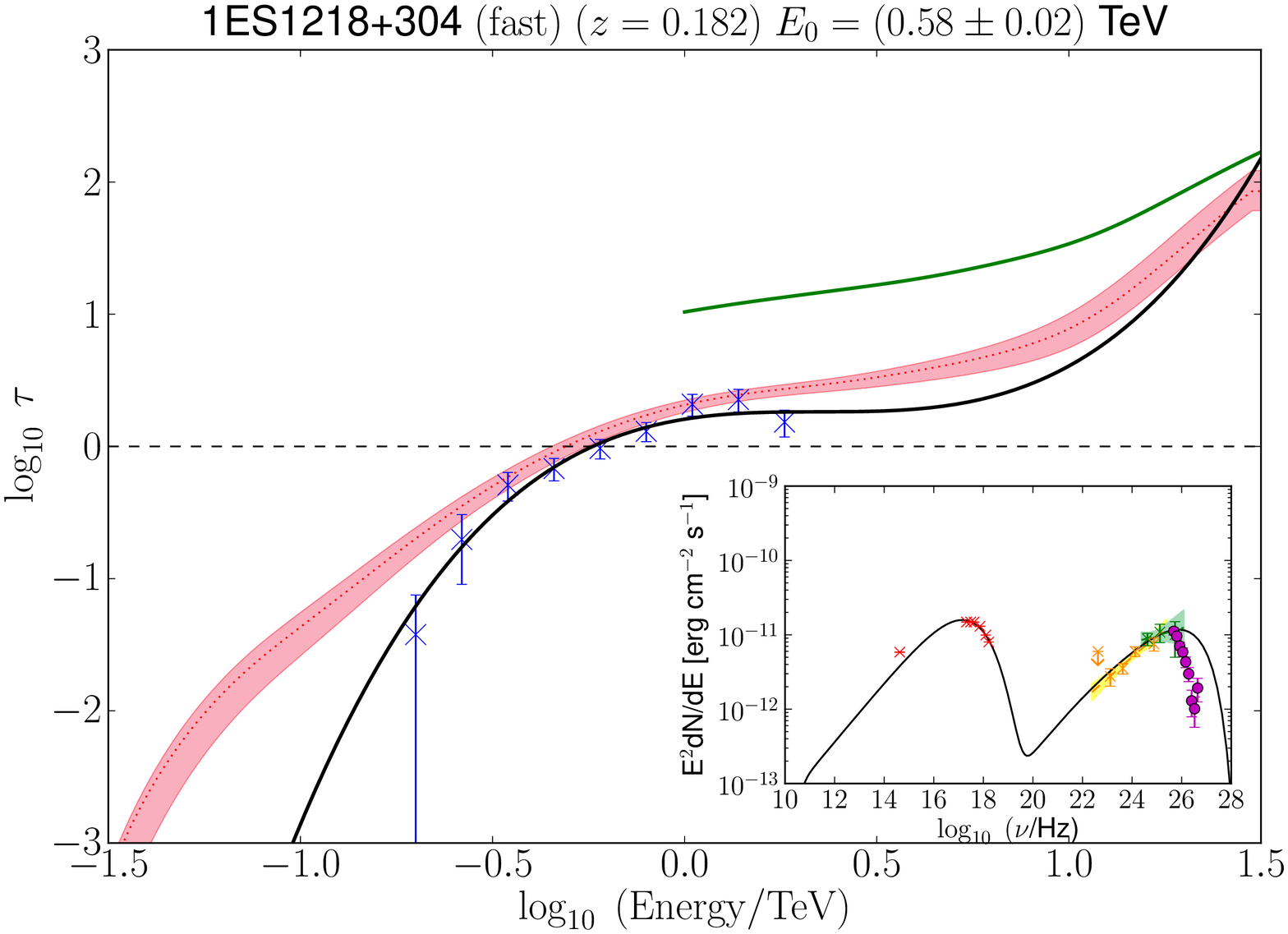}
\includegraphics[trim=1.7cm 0 2cm 0.8cm,clip=True,width=\columnwidth]{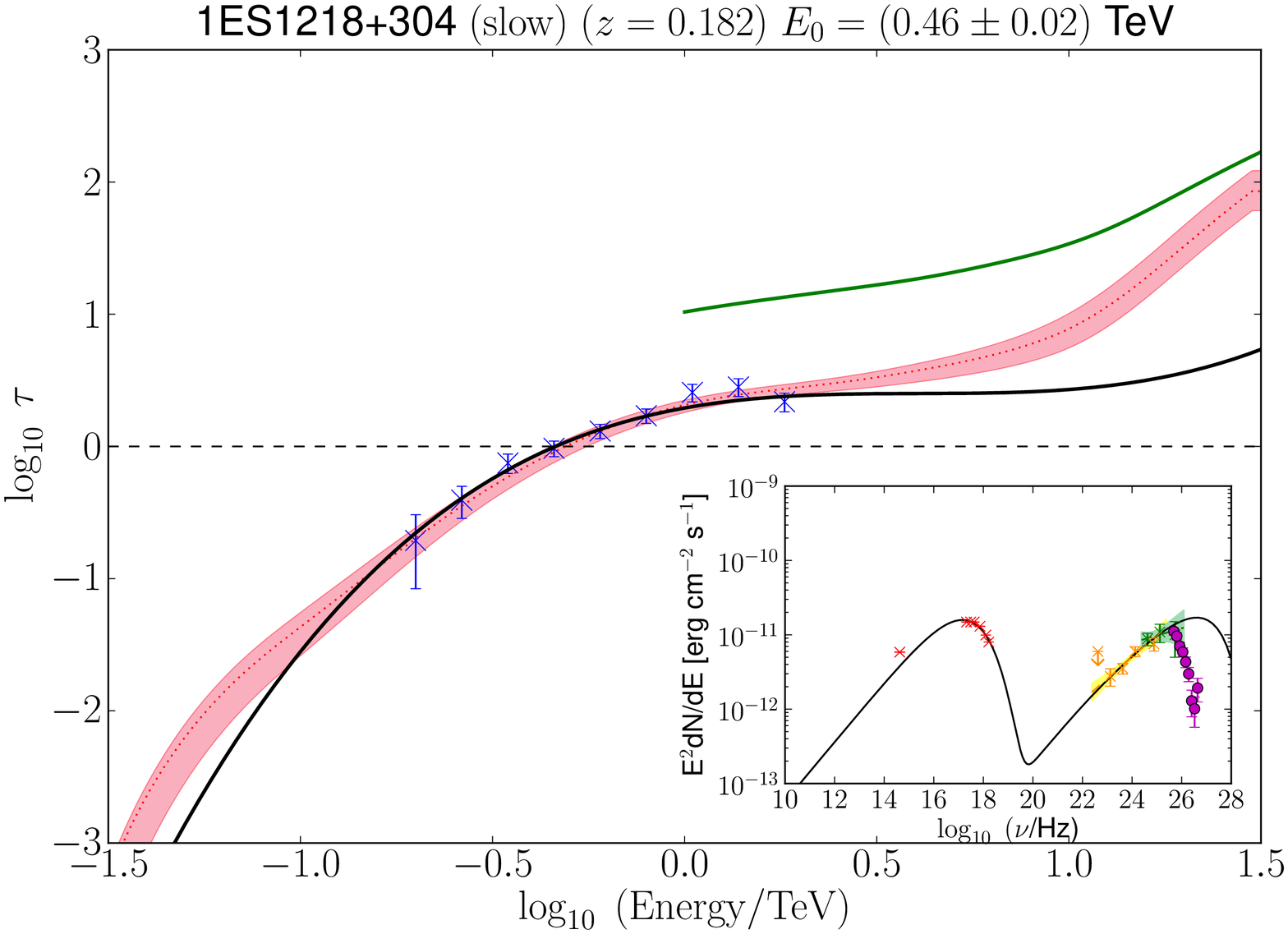}\\
\includegraphics[trim=1.7cm 0 2cm 0.8cm,clip=True,width=\columnwidth]{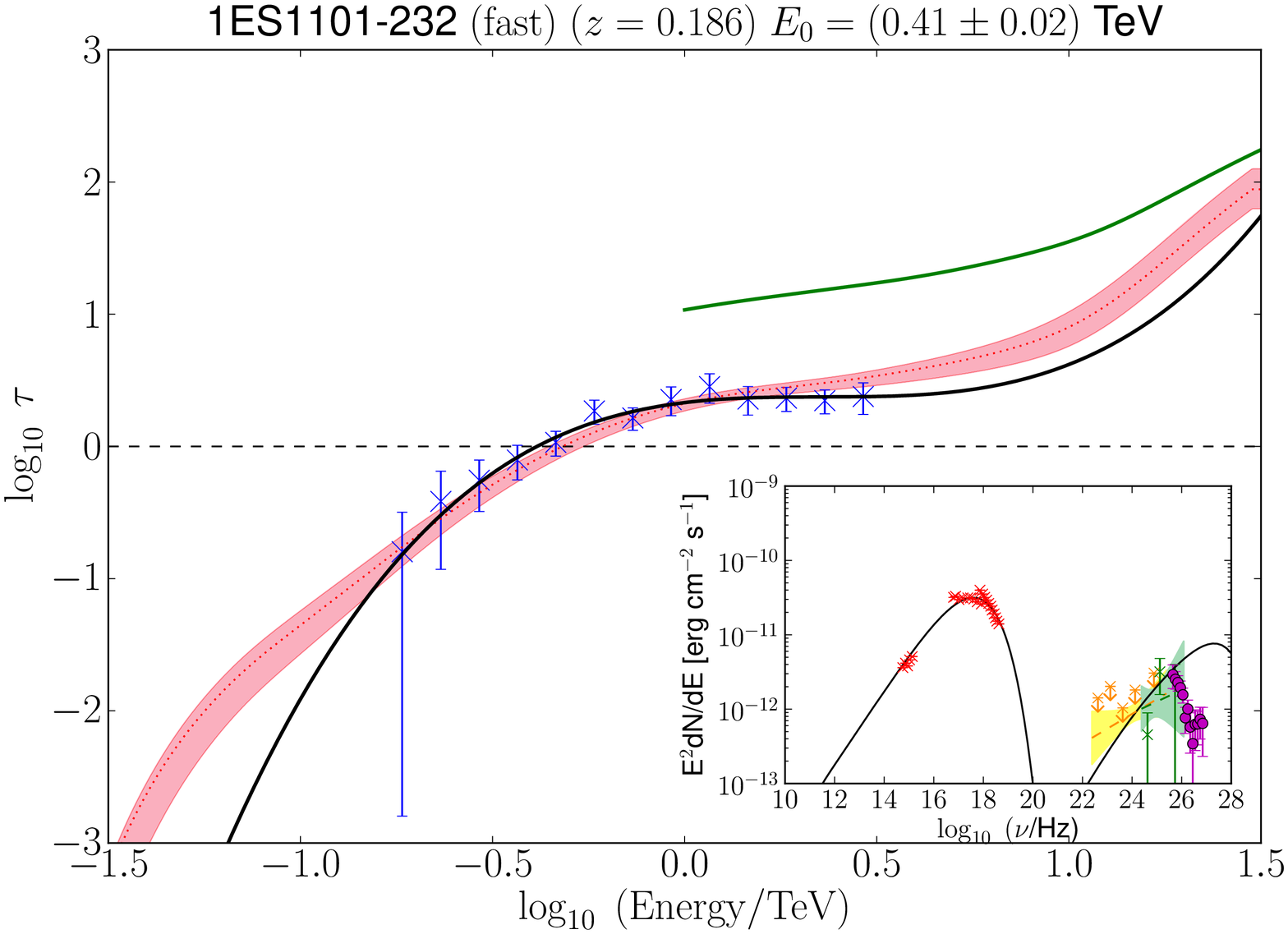}
\includegraphics[trim=1.7cm 0 2cm 0.8cm,clip=True,width=\columnwidth]{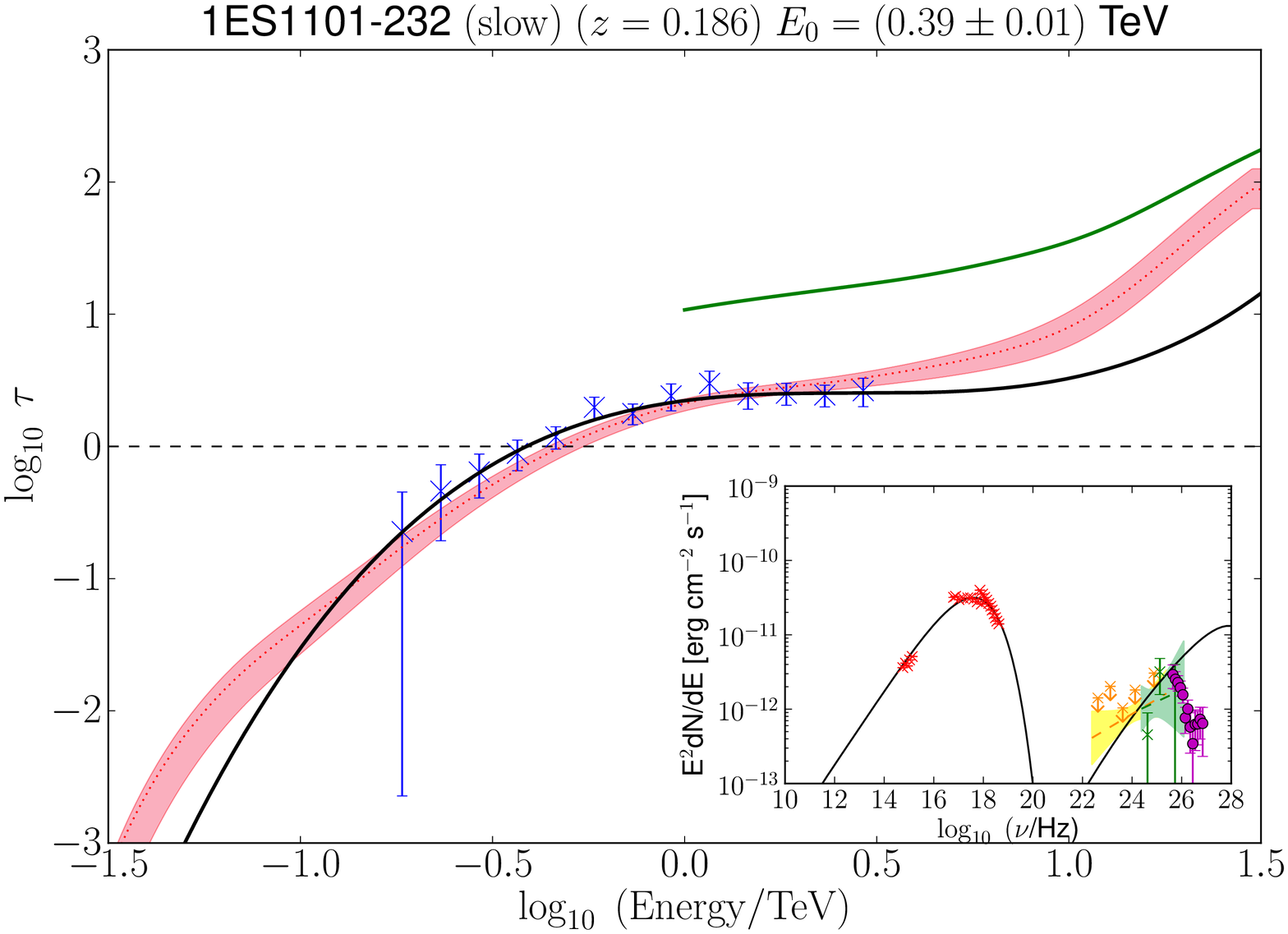}\\
\center{{\bf Figure 1.}---  continued}
\end{figure*}

\begin{figure*}
\includegraphics[trim=1.7cm 0 2cm 0.8cm,clip=True,width=\columnwidth]{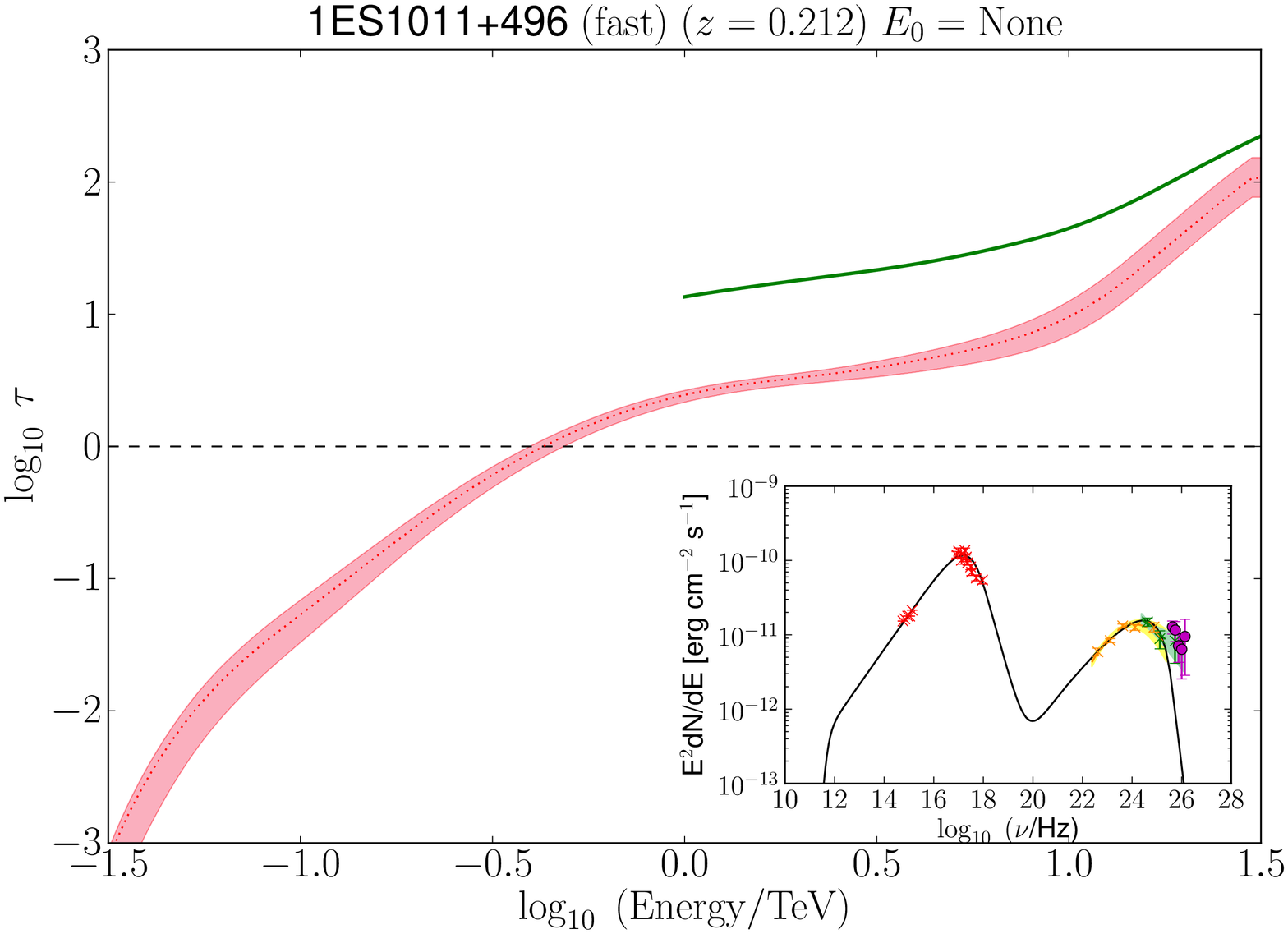}
\includegraphics[trim=1.7cm 0 2cm 0.8cm,clip=True,width=\columnwidth]{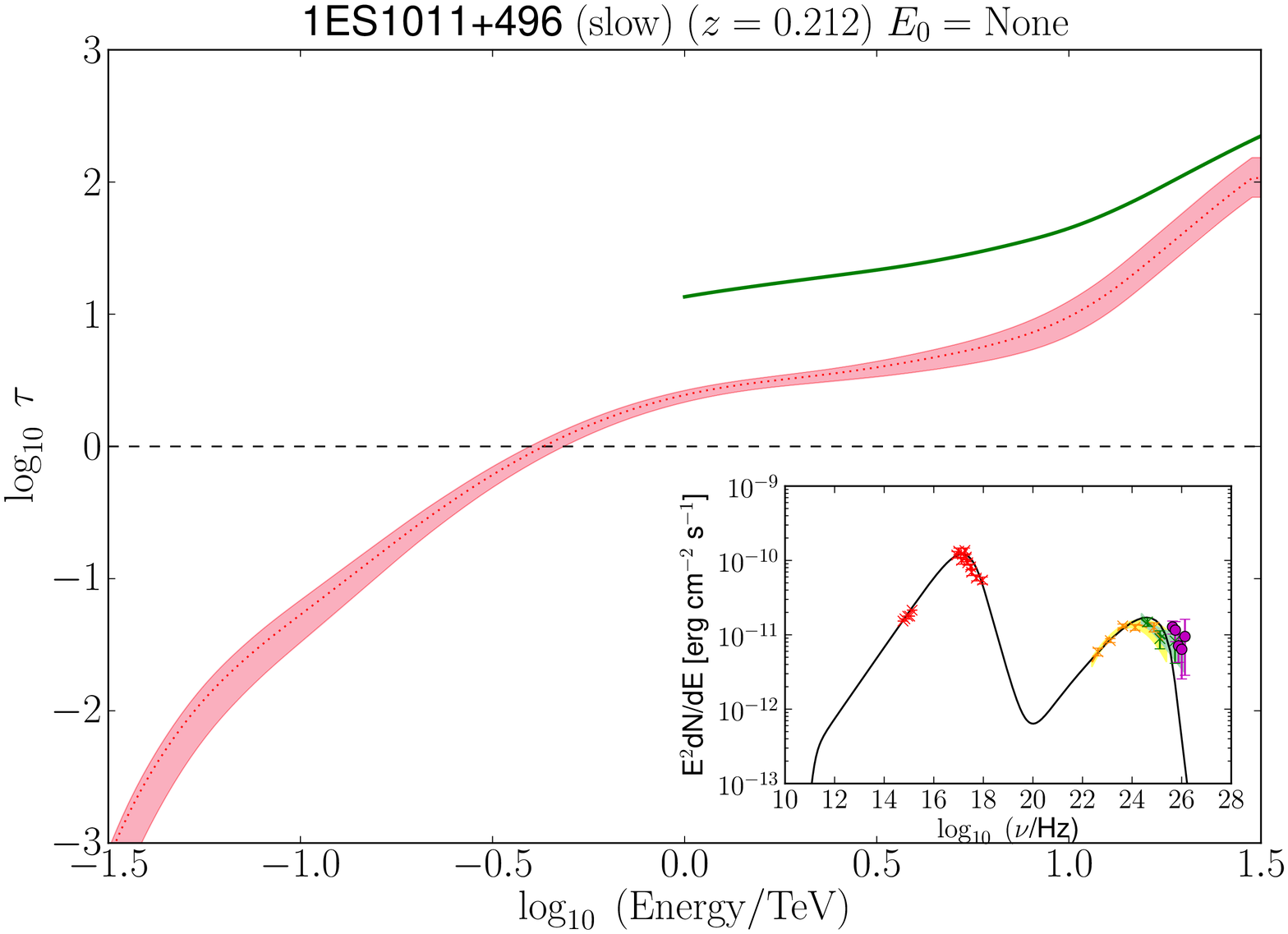}\\
\includegraphics[trim=1.7cm 0 2cm 0.8cm,clip=True,width=\columnwidth]{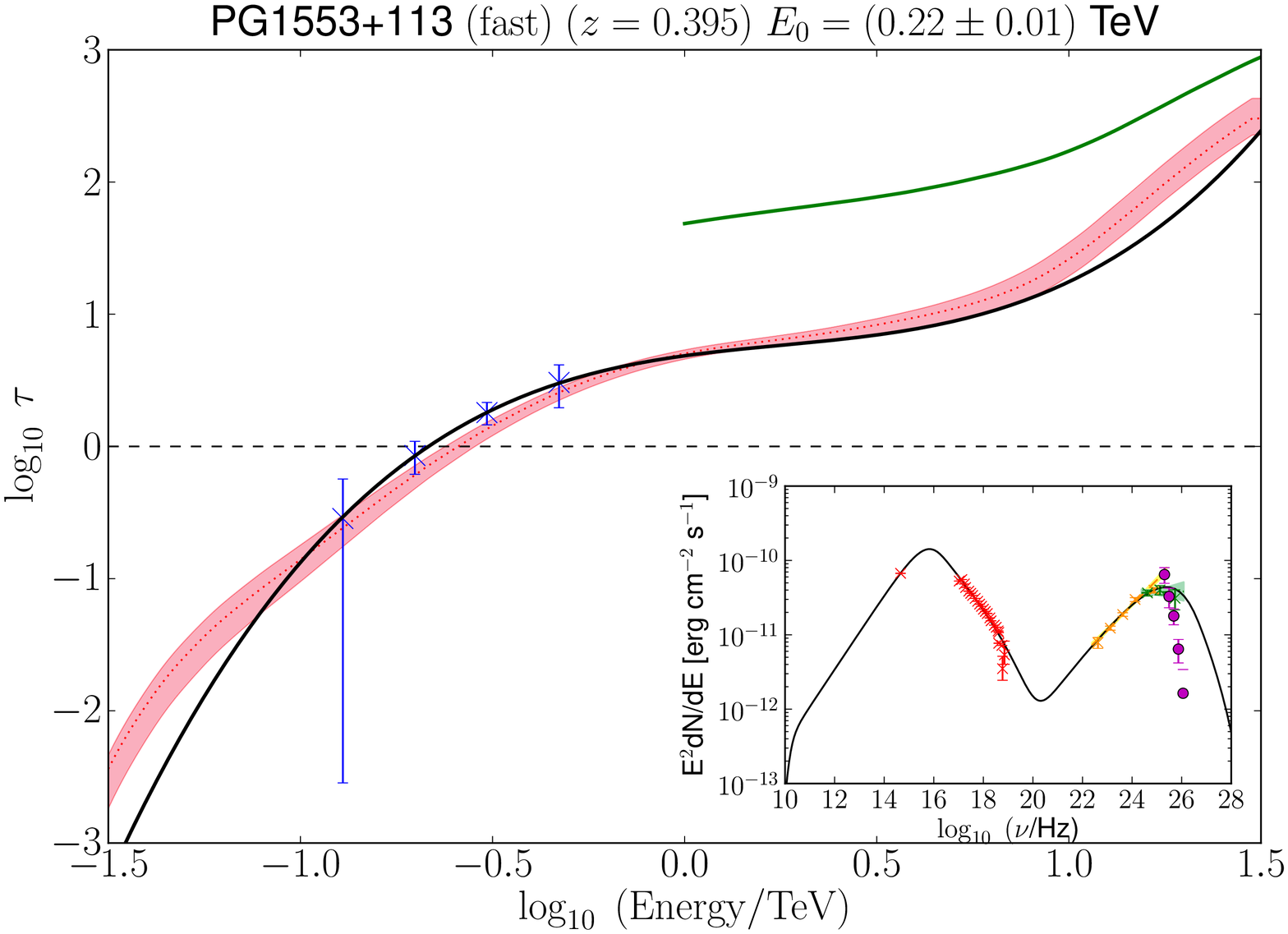}
\includegraphics[trim=1.7cm 0 2cm 0.8cm,clip=True,width=\columnwidth]{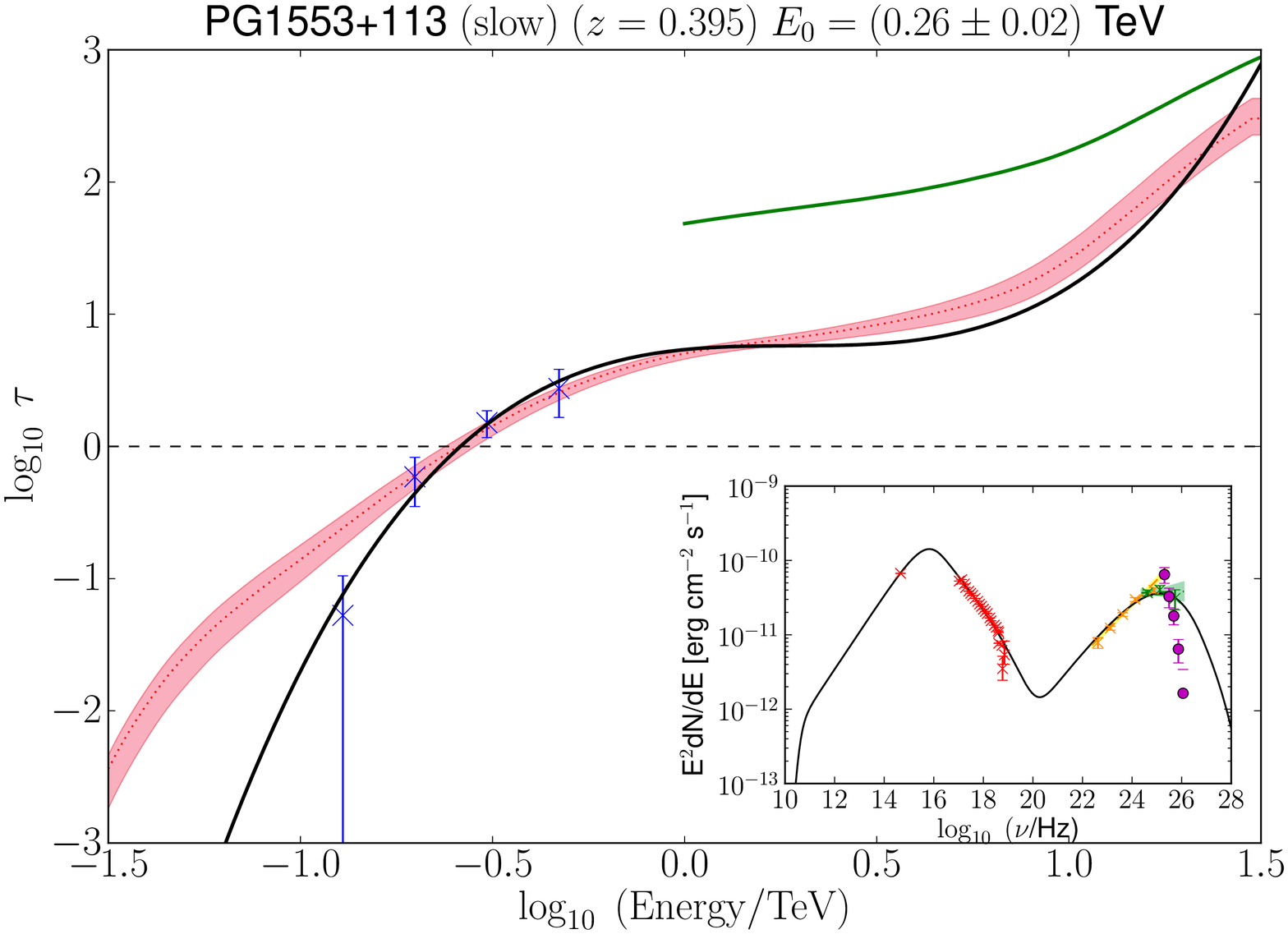}\\
\includegraphics[trim=1.7cm 0 2cm 0.8cm,clip=True,width=\columnwidth]{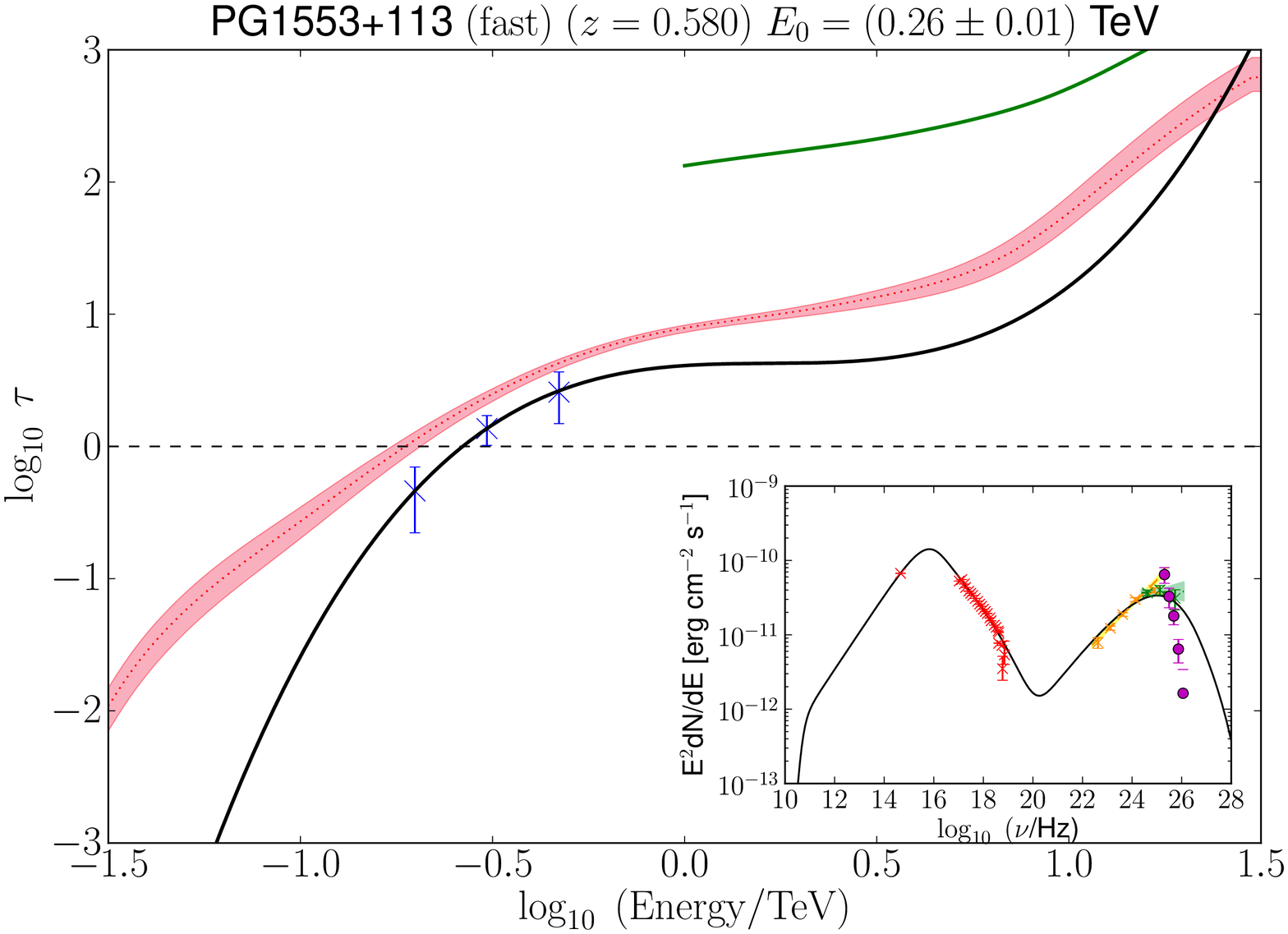}
\includegraphics[trim=1.7cm 0 2cm 0.8cm,clip=True,width=\columnwidth]{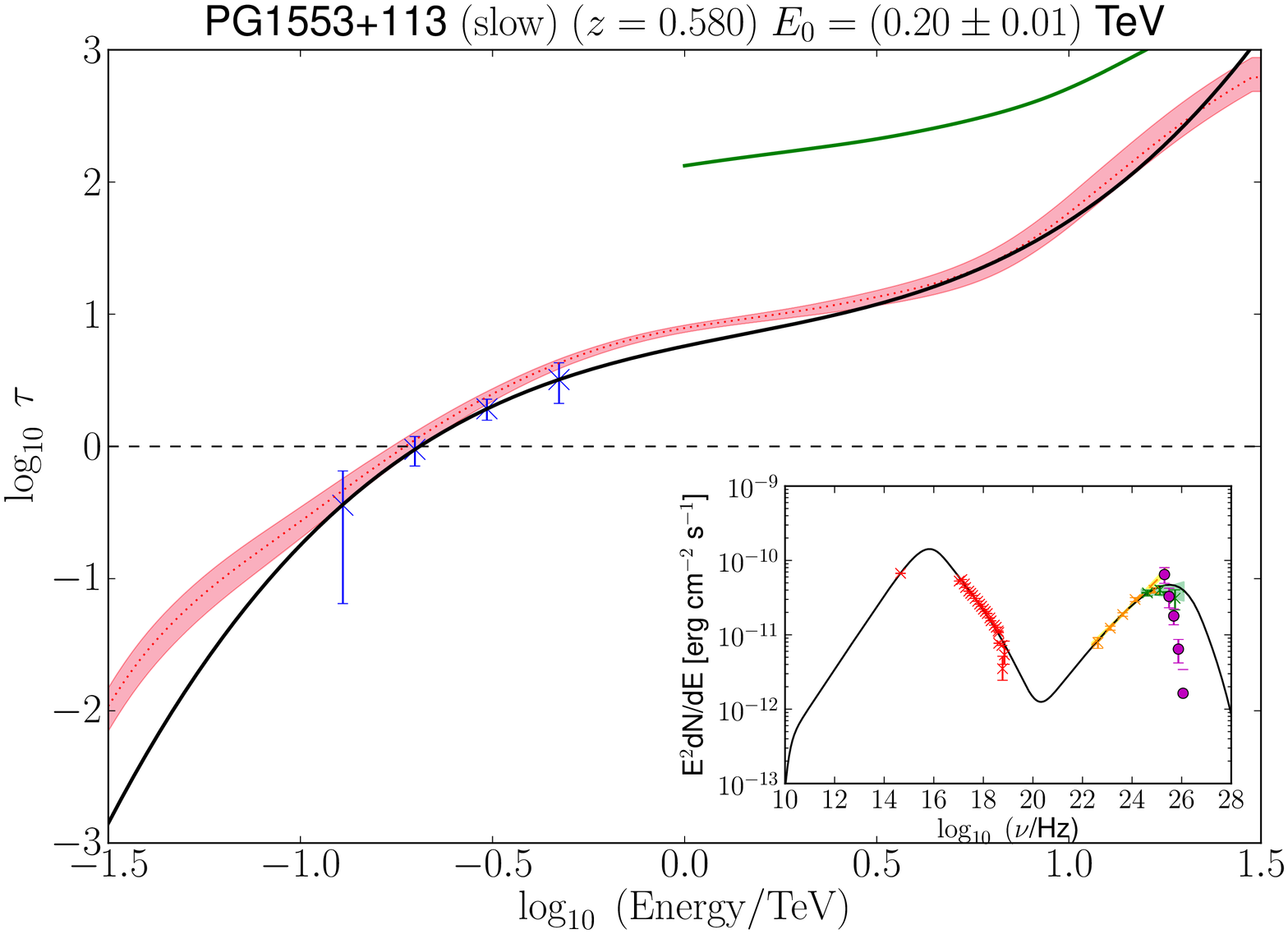}\\
\center{{\bf Figure 1.}---  continued}
\end{figure*}

\begin{figure*}
\includegraphics[trim=1.7cm 0 2cm 0.8cm,clip=True,width=\columnwidth]{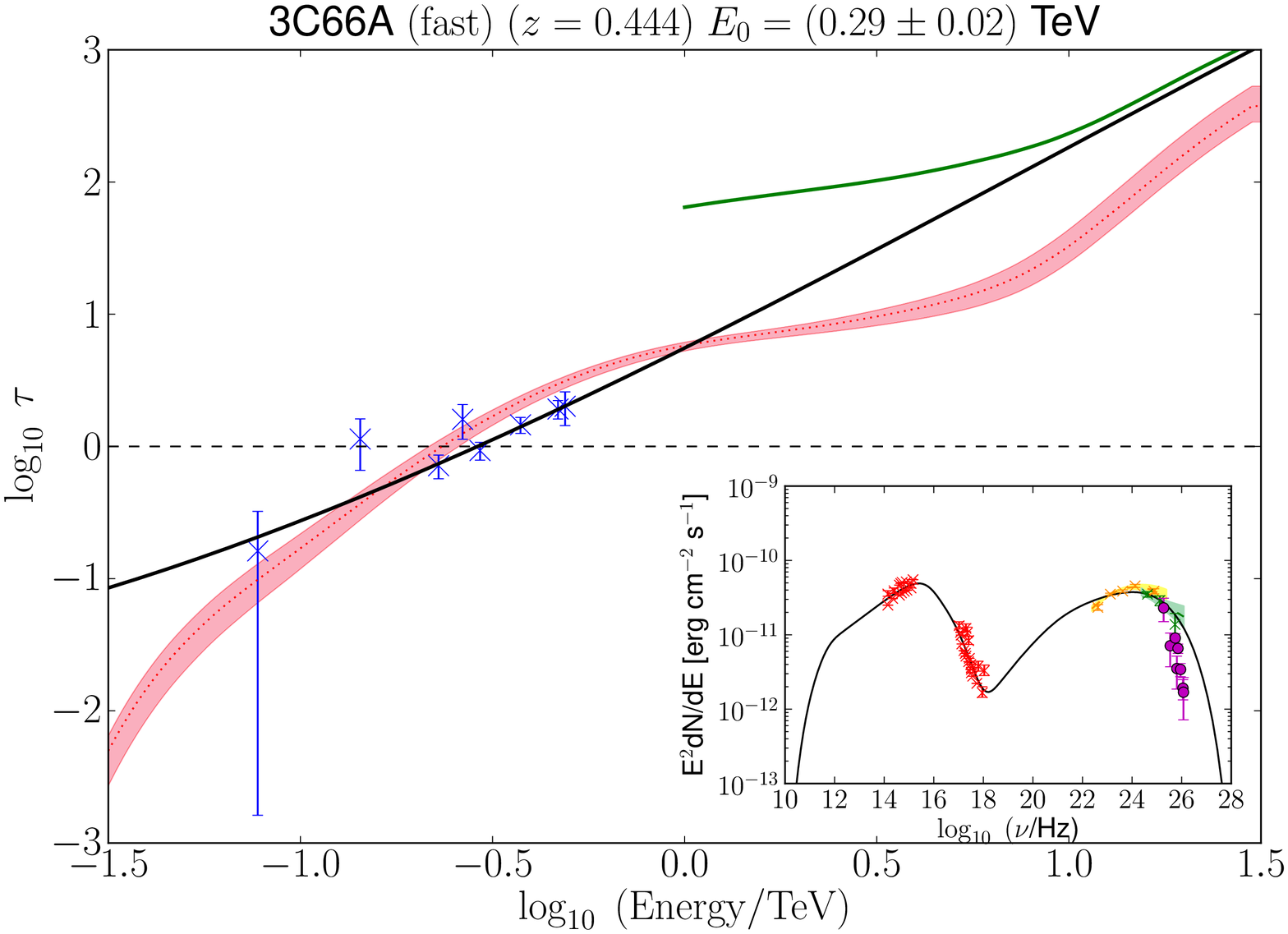}
\includegraphics[trim=1.7cm 0 2cm 0.8cm,clip=True,width=\columnwidth]{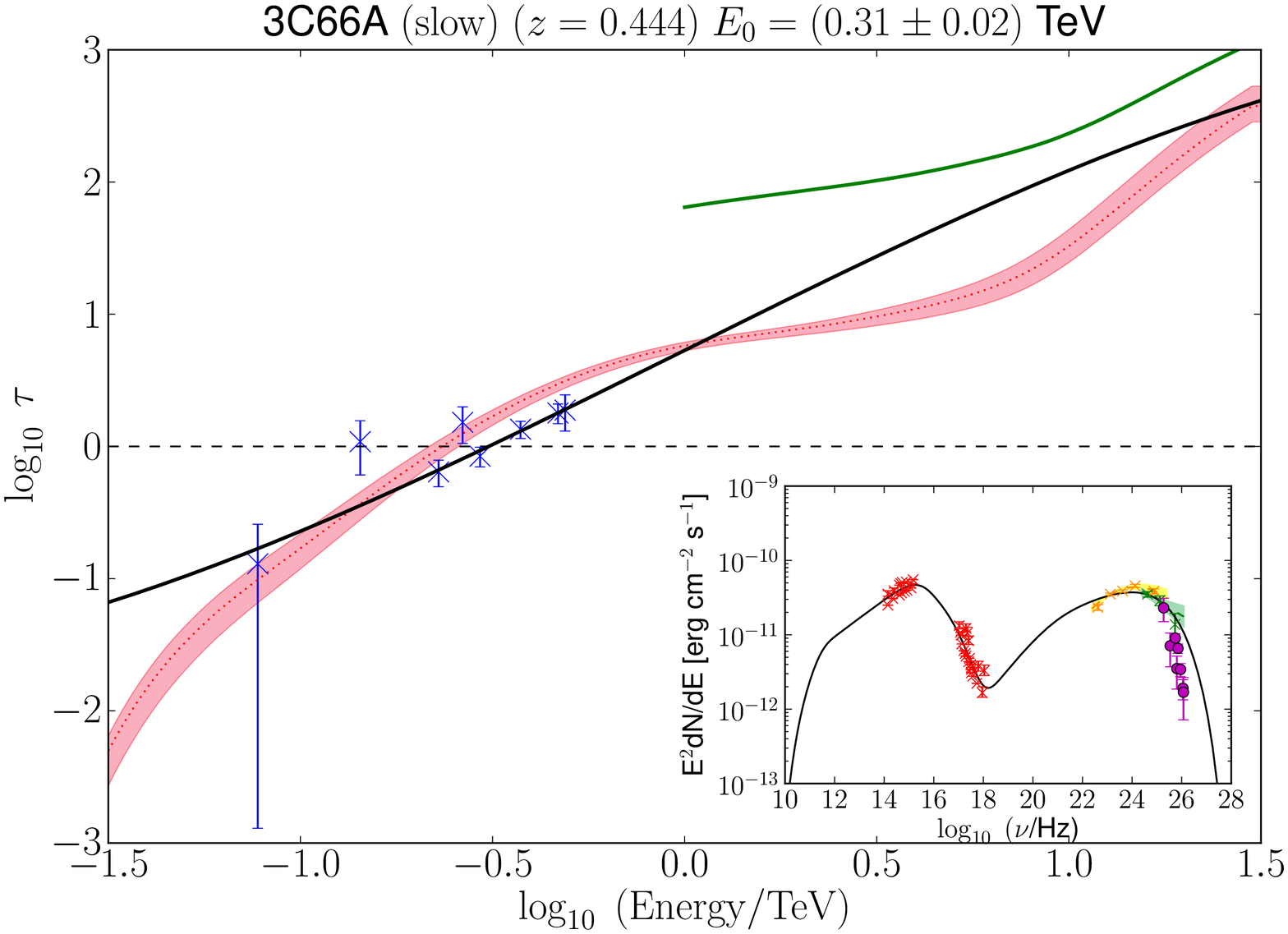}\\
\center{{\bf Figure 1.}---  continued}
\end{figure*}

\subsection{Maximum likelihood polynomial fitting}

We assume that $\log_{10}(\tau)$ (as obtained from equation~\ref{eq:tau}) may be described by a third-order polynomial in $\log_{10}(E)$,

\begin{equation}
\label{eq:poly}
\log_{10}(\tau) = a_{0}+a_{1}\log_{10}(E)+a_{2}\log_{10}^{2}(E)+a_{3} \log_{10}^{3}(E)
\end{equation}

where $E$ is in units of TeV. Lower order polynomials are not sufficient to describe the optical depth. Higher order polynomials introduce too many degrees of freedom and increase the computational time without increasing the precision of our analysis. This shape of the opacity (which is the integral of the EBL spectral intensity and the pair-production cross section; see \eg \citealt{dominguez11a}) is expected in the VHE range for two main reasons. First, the EBL SED must have a smooth shape as a consequence of the galaxy SEDs that produce the EBL and second because of the continuity of the cross-section of the pair-production interactions (\eg \citealt{dwek05}). A maximum likelihood method that scans the parameter space is adopted to compute the likelihood of the estimated data given the different polynomials $\log_{10}(\tau)$ (blue crosses in Figure~\ref{fig:E0fits}) for each blazar. The four parameters of the third order polynomial are explored by studying their probability density distributions. At this point, three physically motivated assumptions are made. First, that $\tau < 1$ at $E=0.03$~TeV; since EBL attenuation is expected to be significantly low at these energies. Second, that $\tau \le$~UL($E,z$), where UL is an upper limit calculated in the present work from the EBL upper limits presented in \citet{mazin07}; in particular $1 \le \tau \le$~UL($z$) at $E=30$~TeV. In \citet{mazin07} two upper limits are presented coming from a realistic and a weaker assumption on the blazar emission that is called \emph{extreme} and that provides the higher upper limits. The \emph{extreme} case is used in our analysis since we want to keep our methodology conservative.

Third, we impose that $\tau$ should increase monotonically with energy. These assumptions will be discussed in more detail in Section~\ref{sec:discussion}. The $E_{0}$ value derived from each blazar, for both slow and fast variability timescales, is then estimated from the most likely polynomial in the four dimensional parameter space (these $E_{0}$ values are given in each inset of Figure~\ref{fig:E0fits} and also in Table~\ref{tab2}). The uncertainty is estimated by using a standard Jackknife analysis (\citealt{wall03}). The final (combined) $E_{0}$ for each blazar is then calculated as the geometric mean value for the two variability timescales\footnote{The geometric mean is chosen over an arithmetic mean since the two $E_{0}$ values for a given source may initially be spread over a wide energy range.} We stress that the uncertainties in this value includes the uncertainties derived from the two SSC modelings (physically the uncertainty in the minimum time variability $t_{v,min}$), which are bracketed by the two different predictions of the VHE fluxes. The lower and upper uncertainties of the combined $E_{0}$ are taken from the $E_{0}-\Delta E_{0}$ and $E_{0}+\Delta E_{0}$ of the state with the lowest and highest $E_{0}$, respectively. We stress that these uncertainties are more conservative than $1\sigma$.

\section{Estimation of the cosmic $\gamma$-ray horizon}
\label{sec:results}
The parameters that describe the synchrotron/SSC models from the fits to the quasi-simultaneous multiwavelength data are listed in Table~\ref{tab1}. We provide two different fits to every blazar bracketing the expected intrinsic VHE fluxes. (These two fits are named slow and fast according to their variability timescale.) The methodology described in the previous section is applied to every blazar in our catalog. As we mentioned in Section~\ref{sec:data}, PG~1553+113 has an uncertain but well constrained redshift (\citealt{danforth10}). Therefore, two different fits are provided for both redshift limits for this blazar.

Figure~\ref{fig:E0fits} shows all the fits for the 15 blazars used in our analysis (two fits per blazar for each minimum variability timescale, except four fits for PG~1553+113 to account for its redshift uncertainty). The fast minimum time variability fits ($t_{v,min}=10^{4}$~s) are shown on the left side of the figure whereas the slow minimum time variability fits ($t_{v,min}=10^{5}$~s) are shown on the right side. Each panel shows the $\log_{10}(\tau)$ data derived from equation~\ref{eq:tau} versus the $\log_{10}$ of the energy in TeV. Figure~\ref{fig:E0fits} shows the upper limits of the optical depth calculated from the EBL upper limits provided by \citet{mazin07} and the most likely polynomials. The $E_{0}$ value is calculated as the energy where $\log_{10}(\tau)=0$ from the most likely polynomial in the maximum likelihood parameter space distribution. For comparison, the estimation of $E_{0}$ calculated from the EBL model based on observations presented by D11 is shown in every panel as a red-dotted line. The uncertainties in the optical depth from the EBL modeling are shown as a red area. Every panel in Figure~\ref{fig:E0fits} also has an inset with the multiwavelength data and the best-fit synchrotron/SSC model ($E^{2}dN/dE$ versus $\log_{10}$ of the frequency in Hz).

\begin{figure*}\center
\includegraphics[width=14cm]{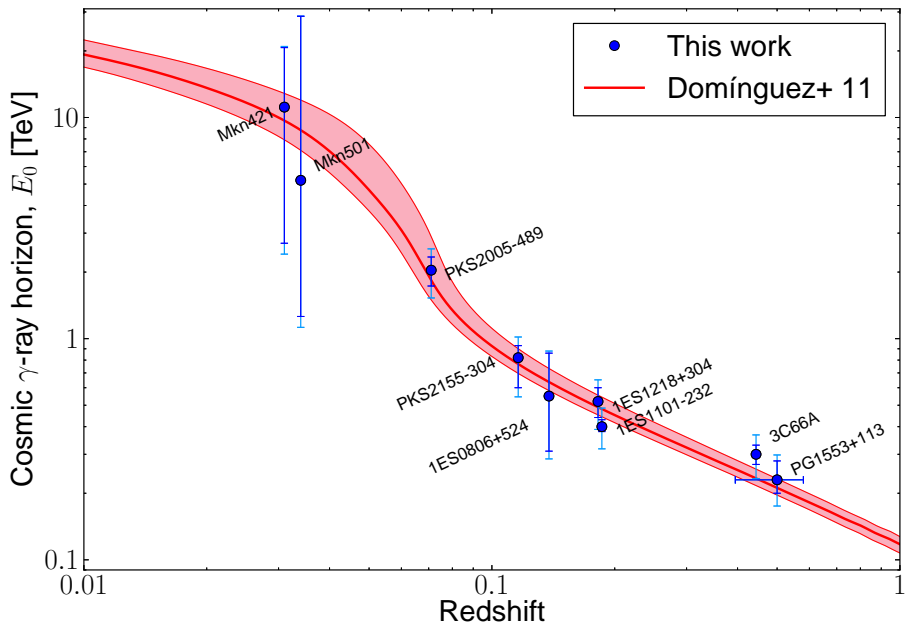}
\caption{Estimation of the CGRH from every blazar in our sample plotted with blue circles. The statistical uncertainties are shown with darker blue lines and the statistical plus 20\% of systematic uncertainties are shown with lighter blue lines. The CGRH calculated from the EBL model described in \citet{dominguez11a} is plotted with a red-thick line. The shaded regions show the uncertainties from the EBL modeling, which were derived from observed data.}
\label{fig:grh}
\end{figure*}

The final $E_{0}$ is then assessed combining the two $E_{0}$ results (slow/fast minimum time variability) from every blazar as the geometric mean value between these two estimates. The statistical uncertainties bracket the two $E_{0}$ values from every fit for every blazar with their uncertainties estimated from the maximum likelihood analysis using the likelihood distributions of the polynomial parameters. In our analysis, we also consider the systematic uncertainties in the LAT measurements. Their effect is estimated by artificially hardening and softening the overall SEDs of PKS~2005$-$489 (a blazar with low statistical uncertainties). First, the fluxes of the three lowest-energy LAT data for PKS~2005$-$489 are decreased by 10\% (which is the typical \emph{Fermi}-LAT systematic uncertainty of the effective area, \citealt{ackermann12a}) and the three highest-energy LAT bins are increased in flux by 10\%. The overall SED fit is done with these new points, and the energy where $\tau=1$ is estimated, using the procedure described in section~\ref{sec:methodology}. This procedure is repeated by increasing by 10\% the three lowest-energy LAT data of PKS~2005$-$489 whereas decreasing by 10\% the three highest-energy LAT data. This allows us to estimate an average systematic uncertainty of 20\% in the energy where $\tau=1$ for PKS~2005$-$489. We thus assume the systematic uncertainty from the uncertainty in the LAT is 20\% for all sources. The observed CGRH is shown in Figure~\ref{fig:grh} with blue circles, the statistical uncertainties are shown with darker blue lines, and the statistical plus systematic uncertainties (added in quadrature) with lighter blue. A completely independent estimation of the CGRH from the EBL model described in D11 is also shown with its uncertainties, which are thoroughly discussed in D11. The uncertainties in the EBL modeling are larger in the far-IR region for the reasons discussed in D11. This leads to the larger uncertainties in the estimation of the CGRH from the EBL modeling at the lower redshifts. The reason is that this is the EBL region that mainly interacts with the higher-energy VHE photons that lead to determination of the CGRH in that redshift range.

Our methodology offers more information on the optical depth than just the CGRH. Therefore, the same procedure followed to calculate the CGRH is applied to calculate the energies at which the optical depth is equal to 0.5, 2, and 3 (shown in Figure~\ref{fig:morecgrh} with blue squares, green triangles, and magenta diamonds, respectively). The energies for those optical depths are plotted from the D11 model with their uncertainties as well (the same colors are used for each modeled optical depth as for the data). The uncertainties in these estimates are significantly larger since, in general, they are given by the extrapolation of the most likely polynomial outside the energy range of the Cherenkov detections.

\section{Discussion}
\label{sec:discussion}
In this work, we present an estimation of the CGRH based on a multiwavelength compilation of blazars that includes the most recent \emph{Fermi}-LAT data. We stress that our estimation of the CGRH is derived with only a few physically-motivated constraints. These results represent a major improvement with respect to previous works. These previous works provide only lower limits for the CGRH such as the EBL-model dependent limits estimated by \citet{albert08} (that are based on a modified parameterization of the EBL models presented by \citealt{kneiske02}). Other CGRH limits are presented by \citet{abdo10b} using only \emph{Fermi}-LAT observations.

The \emph{Fermi}-LAT hard-source catalog (Ackermann et al., in preparation) is included in our analysis. The inclusion of this data set in our multiwavelength blazar catalog is essential for the right estimation of the CGRH since these measurements help to resolve the shape of the inverse Compton peak.

The optical depth is calculated using equation~\ref{eq:tau}, which describes the ratio between the intrinsic flux from the synchrotron/SSC models and the observed flux by IACTs. Then, these data are fitted to polynomials of third order imposing some constraints. We also require an increasing and monotonic behavior of the polynomials. Polynomials of order lower than three would not reproduce the expected optical depth shape while larger order polynomials would introduce unnecessary parameters into the fits. The constraints are all physically motivated and EBL-model independent. As we said before, the first condition is that $\tau \le 1$ at $E=0.03$~TeV, which means that the attenuation is rather weak at those low energies.

The second constraint is that $1\le \tau \le$~UL($z$) at $E=30$~TeV, where UL are the opacities calculated from the EBL upper limit in the local Universe found in \citet{mazin07}. The upper limits of their so-called \emph{extreme} case are used in our analysis. This \emph{extreme} case represents the least constraining assumption on the blazar spectra since it allows us a wider range of spectral indices (\ie this results in a rather conservative hypothesis for our analysis). For this same reason, we prefer to use as conservative upper limits the results by \citet{mazin07} rather than the newer results by \citet{meyer12} that are based on a more constraining spectral condition. The EBL evolution is expected to affect the optical depth calculated at higher redshifts. To account for this effect we evolve conservatively the EBL upper limits at all wavelengths as $(1+z)^{5}$ (in the co-moving frame) when calculating the optical depths from these EBL limits from \citet{mazin07}. We note that this is a robust limit given the fact that the maximum evolution (which is dependent on the wavelength) is $(1+z)^{2.5}$ in a realistic model such as D11 for $0\le z \le 0.6$ (the redshift range of our blazar catalog).

The third constraint that we apply for our fits is to require only monotonically increasing functions for $\log_{10}(\tau)$ as a function of $\log_{10}(E)$. This condition is also expected for any realistic EBL spectral intensity, which comes from galaxy emission, given the increasing behavior of the pair-production interaction with energy. Interestingly, we see in Figure~\ref{fig:E0fits} that in most cases the IACT observations are indeed detecting the flux decrement given by the CGRH feature (\ie the Cherenkov observations span from negative to positive values of $\log_{10}(\tau)$).

\begin{figure*}\center
\includegraphics[width=14cm]{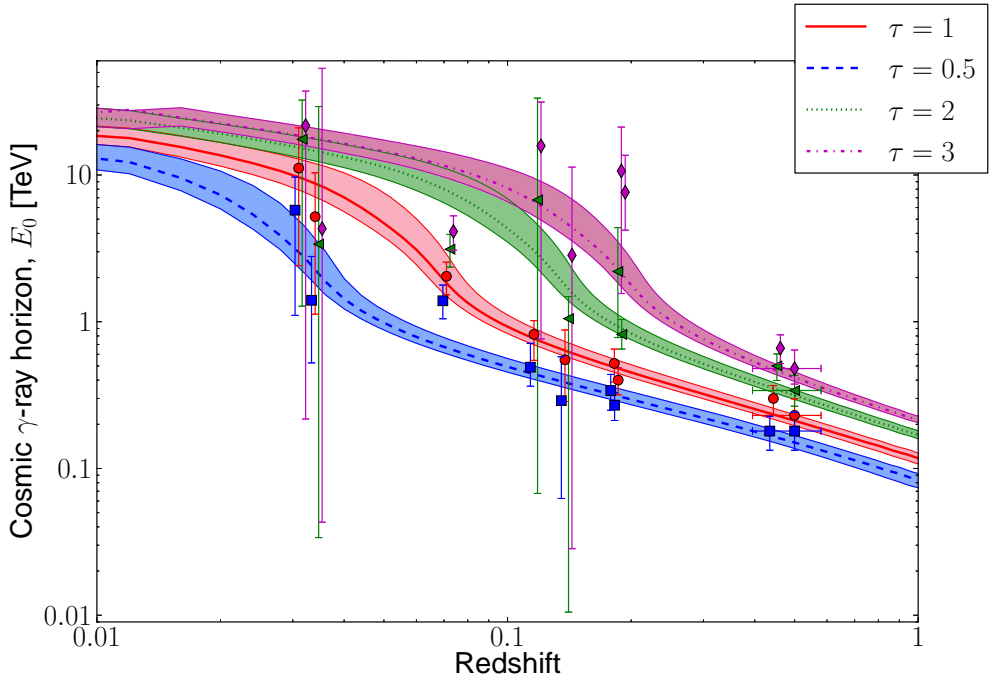}
\caption{Energy values at which the optical depth is 0.5 (blue squares), 1 (red circles), 2 (green triangles) and 3 (magenta diamonds) from both the blazars presented in the current analysis and the EBL model described in \citet{dominguez11a}. The shaded regions show the uncertainties from the EBL modeling (the same colors are used for each modeled optical depths as for the data), which were derived from observed data. The different data for a given blazar are slightly shifted in the $x$-axis for clarity.}
\label{fig:morecgrh}
\end{figure*}

We find that the CGRH derived from 9 out of 11 blazars where our maximum likelihood methodology can be applied, is compatible with the expected value from the D11 model. The estimations from other EBL models such as \citet{franceschini08}, \citet{finke10} (model $C$), and \citet{somerville12} are in agreement within uncertainties with the EBL model by D11. We note that the fit of 1ES~1101$-$232 has only one degree of freedom, see Table~\ref{tab1}. The uncertainties of the two lowest redshift blazars (Mkn~501 and Mkn~421) are systematically higher because the optical depth for these cases becomes unity at energies larger than the energies observed by the Cherenkov telescopes. Therefore, in these cases $\tau=1$ is given by an extrapolation of the polynomials rather than an interpolation between observed energies (see Fig.~\ref{fig:E0fits}) leading to greater uncertainty. For the case of 1ES~2344+514 with fast flux variability timescale, a value of $E_{0}$ in agreement with the estimation by the D11 EBL model is derived. However, for this case the uncertainties are larger than $E_{0}$ and therefore no useful constraint can be derived. For the case of 1ES~2344+514 with slow flux variability timescale, the SSC predicted flux is lower than the flux given IACT data. For H~1426+428, both flux variability timescales give uncertainties in the measurement of $E_{0}$ larger than $E_{0}$ and therefore no constraint can be derived. In both cases the synchrotron/SSC model does not seem to correctly fit the multiwavelength data. Our maximum likelihood procedure cannot be applied to any flux state on four blazars (1ES~1959+650, W~Comae, H~2356$-$309 and 1ES~1011+496). There are different explanations for this fact. Some blazars have shown flux variability on the scale of minutes (\eg \citealt{aharonian07,albert08,aleksic11b,arlen13}) and the IACTs tend to detect the sources in higher-flux states. In most cases, the LAT data are not simultaneous with the IACT and other multiwavelength data. We have tried to alleviate this problem by choosing SEDs that are based on a low, non flaring state, where the variability seems to be small. In this way the effects of variability from epoch to epoch have been minimized. We compare the long-term light curves in X-rays using the quick-look results from the All Sky Monitor (ASM) aboard the Rossi X-Ray Timing Explorer\footnote{http://xte.mit.edu/ASM\_lc.html} (RXTE) with the time range of the IACT observation for those four blazars where our maximum likelihood procedure could not be applied (see the Appendix for more details). Clearly 1ES~1011+496 was indeed detected by the IACTs in flaring states. The situation for 1ES~1959+650 is not clear. And the light curve of the H~2356$-$309 observation was rather irregular. We could not find X-ray data for W~Comae in the ASM database but this source was clearly detected in TeV in a strong flare (\citealt{acciari08}).

The synchrotron/SSC model is the standard model for fitting high-peaked TeV BL Lac objects, and does seem to provide a good fit to their broadband SEDs (\eg \citealt{zhang12}). However, there are some alternatives. High-peaked BL Lac objects are not thought to have a significant contribution to the $\gamma$-ray flux from scattering external photon sources, but there are some exceptions, such as the eponymous BL Lac (\citealt{abdo11c}). It has also been suggested that for some sources, a lepto-hadronic model provides a better fit, such as 1ES~0414+009 (\citealt{aliu12}). Non-variable TeV emission unrelated to the rest of the broadband SED could originate from Compton-scattering of the cosmic microwave background (CMB) by an extended jet (\citealt{boettcher08}), which would certainly complicate their modeling. Another way of creating non-variable TeV emission unrelated to the SED, which would also avoid much of the EBL attenuation, would be if the AGN produces a significant number of cosmic rays, which during propagation interact with the CMB and EBL to produce the observed $\gamma$-rays (\citealt{essey10,essey11}). Finally, even if the synchrotron/SSC model is valid for the TeV blazars considered here, it is possible that the electron-positron pairs created by the VHE $\gamma$-ray interactions with EBL photons can Compton-scatter the CMB, producing $\gamma$-rays observable by the LAT (\citealt{neronov10,tavecchio10,taylor11,dermer11,vovk12}). This would complicate the modeling process, since it would add other, poorly-constrained parameters (\citealt{tavecchio11}). Nonetheless, the simple synchrotron/SSC is a very attractive model, due to its success at fitting a large number of objects, and its relatively small number of free parameters. The existence of axion-like particles could also allow the $\gamma$-rays to avoid the photoabsorption process, changing the expected VHE spectrum (\citealt{sanchezconde09,dominguez11d}). A detailed broadband modeling including all these non-standard considerations is out of the scope of this work.

In summary, we built a catalog of 15 blazars from Zhang et al. (2012) by requiring a good multiwavelength spectral coverage and TeV detections. After fitting a synchrotron/SSC model to each source, there were four blazars where the TeV detection was at higher fluxes than the flux extrapolation from the models. As described above, the long-term X-ray data from RXTE were checked on the dates of the TeV observations for these blazars showing a hint that H~2356$-$309 and 1ES~1011+496 were flaring in X-ray and therefore probably in the TeV range as well. The situation is not clear for 1ES~1959+650. However, as described above potential problems with the synchrotron/SSC model cannot be ruled out. There are two other blazars (H~1426+428, 1ES~2344+514) for which the uncertainties that we derive for $E_{0}$ are too high to set any constrain. This is due to large uncertainties in the TeV measurements and/or the low number of TeV spectral points, which do not allow a reliable $E_{0}$ estimation.

The agreement between the CGRH observation presented in this work and the expected values from D11 indicates that these possibilities described above might not be relevant for many blazars. Furthermore, Figure~\ref{fig:morecgrh} gives more information on the optical-depth shape derived from our methodology. This figure shows that the energies at which the optical depths are 0.5, 1 (the standard definition of CGRH), 2 and 3 are still compatible with the D11 model. The uncertainties are generally higher at $\tau$ different from 1 due to the fact that the Cherenkov detections do not span the energy range needed in order to derive a better estimation of those energies. As seen in Figure~\ref{fig:E0fits}, the polynomials cut the horizontal lines of constant optical depths generally in wider energy ranges for $\tau$ values different from 1 (\ie $\log_{10}(\tau)=0$). The agreement between the observed CGRH and the expected CGRH from D11 is also consistent with 3C~66A being located at a redshift slightly lower than $z\sim 0.444$\footnote{After our submission, \citet{furniss13} spectroscopically confirmed that 3C~66A is located at $0.3347<z<0.41$ with 99\% confidence.} and PG~1553+113 at $0.395\leq z\leq 0.58$. An independent confirmation of these redshifts will give support to both the current EBL knowledge and our methodology to derive the CGRH. By assuming an EBL model, it is possible to estimate redshifts using EBL attenuation in a realistic way considering the overall SED of the blazars (see the proceeding by \citealt{mankuzhiyil11}). However, we note that some previous estimation of the redshift is necessary in order to fit the synchrotron/SSC models (\citealt{abdo10a,abdo11a}).


\citet{orr11} claimed that recent EBL models such as D11 are incompatible with IACTs observations at more than $3\sigma$. They based their conclusions in an analysis of $\sim 12$ blazars using two different methods that they call \emph{Method~1} and \emph{Method~2}. Their more constraining results are derived from Method~2 (see section~3.2 in \citealt{orr11}). This approach is based on the expected difference between spectral indexes when the VHE spectrum is fitted by a broken power law. This spectral difference is attributed to EBL attenuation, setting limits on the intensity of the local EBL. We consider their results inconclusive. Their Method~2 relies on the assumption that the VHE spectra may be well fitted by broken power laws. We performed F-tests on all the fits of their blazar sample to test whether broken power laws (fitted by two different spectral indexes) could be actually considered better fits to the observed VHE spectra than simple power laws (fitted by a single spectral index)\footnote{An F-test gives the probability that the reduction in the $\chi^{2}$ of the fit due to the inclusion of an additional parameter in the model exceeds the value that can be attributed to random fluctuations in the data; see \citealt{dwek05}.} The F-tests performed for every one of their spectra show that for only 2 out of 12 cases (RGB~J0152+017 and 1ES~1101-232) the spectra can be considered fitted better by broken power laws than simple power laws. Their results from Method~1 (see section~3.1 in \citealt{orr11}) are inconclusive as well. This method relies on the assumption that the VHE intrinsic spectrum is described by a power-law extrapolation of the \emph{Fermi}-LAT data points. As we see in the synchrotron/SSC fits of Figure~\ref{fig:E0fits}, this is not a realistic assumption due to the shape of the inverse Compton peak. Furthermore, we show in the present work that a more sophisticated SSC-based analysis is compatible with the current EBL knowledge.

Some authors have treated the \emph{Fermi}-LAT spectrum, extrapolated into the VHE regime, as an upper limit on the intrinsic spectrum, and used this to compute upper limits on $\tau(E,z)$ (\citealt{georganopoulos10,meyer12}). This provides only upper limits on $\tau(E,z)$ rather than measurements, as we derive here. However, their techniques involve fewer assumptions about the blazar emission model and variability of the SED (see the discussion above). Thus, the two techniques are complementary.

Recently, \citet{ackermann12b} and \citet{abramowski13} claimed the detection of an imprint of the EBL in the blazar spectra. \citet{ackermann12b} base their analysis on \emph{Fermi}-LAT data from $z\sim 0.2$ to 1.6, whereas \citet{abramowski13} use H.E.S.S. data from blazars located at $z\sim 0.1$. These works do not give any results in terms of the CGRH but we are able to estimate it from their results. We find that the results presented in our analysis are compatible with the results from these two independent works, which supports our conclusions.

From our results, we can conclude that the EBL data from direct detection by \citet{cambresy01}, \citet{matsumoto05}, and \citet{bernstein07} are likely contaminated by zodiacal light. This possibility has indirectly been proposed previously by several authors such as \citet{aharonian06}, \citet{mazin07}, and \citet{albert08} using EBL upper limits but we confirm these results using a more robust approach.

\section{Summary} \label{sec:summary}

The CGRH horizon is detected in this work for the first time from a multiwavelength sample of blazars that includes the more recent \emph{Fermi}-LAT data. Only a few general and physically motivated constraints on the EBL were necessary. As we see from our analysis the observational estimation of the CGRH is compatible within uncertainties with the derivation by the observational EBL model described by \citet{dominguez11a}, which is in agreement with the observational EBL model by \citet{franceschini08} and the theoretical methodology followed by \citet{somerville12} and \citet{gilmore12}. All these EBL models are realistic representations of the current knowledge of the EBL (\citealt{dominguez11b,primack11,dominguez12}). We have shown the ability of our methodology to study the opacity of the Universe at different redshifts and to infer distances of blazars with unknown redshifts. Our methodology is sensitive to the total EBL, which includes light even from the faintest and most distant galaxies in the Universe. This will allow us to set limits on the faint-end slope of the evolving galaxy luminosity function, which still remains controversial (see \eg \citealt{reddy09}). The detection of the CGRH presented here will provide an independent test for cosmology and for the estimation of the cosmological parameters that will be presented in Dom\'inguez \& Prada (submitted).

Our technique will benefit in the future with the improved statistics that \emph{Fermi} will provide. The future Cherenkov Telescope Array is expected to provide VHE spectra with a better energy resolution, observed up to higher energies, and increase considerably the number of sources, which indeed will improve the CGRH determination. These prospects together with the increasing number of simultaneous multiwavelength observational campaigns (\eg \citealt{abdo11a,abdo11b,abdo11d}) are promising for a better estimation of the optical depths due to EBL attenuation using our methodology and for the estimation of the CGRH to $z>0.5$.

\section*{Acknowledgments}
The authors thank Marco Ajello, M.~A. S\'anchez-Conde, Seth Digel and David Williams for fruitful discussions. We thank Soebur Razzaque for useful comments on the draft and the anonymous referee for improving the manuscript. We acknowledge the support of the Spanish MICINN's Consolider-Ingenio 2010 Programme under grant MultiDark CSD2009-00064. JRP acknowledges the support from a \emph{Fermi} Guest Investigator grant and also from the NASA ATP grants NNX07AGG4G, NSF-AST-1010033 and NSF-AST-0607712.

The \textit{Fermi} LAT Collaboration acknowledges generous ongoing support from a number of agencies and institutes that have supported both the development and the operation of the LAT as well as scientific data analysis. These include the National Aeronautics and Space Administration and the Department of Energy in the United States, the Commissariat \`a l'Energie Atomique and the Centre National de la Recherche Scientifique / Institut National de Physique Nucl\'eaire et de Physique des Particules in France, the Agenzia Spaziale Italiana and the Istituto Nazionale di Fisica Nucleare in Italy, the Ministry of Education, Culture, Sports, Science and Technology (MEXT), High Energy Accelerator Research Organization (KEK) and Japan Aerospace Exploration Agency (JAXA) in Japan, and the K.~A.~Wallenberg Foundation, the Swedish Research Council and the Swedish National Space Board in Sweden.

Additional support for science analysis during the operations phase is gratefully acknowledged from the Istituto Nazionale di Astrofisica in Italy and the Centre National d'\'Etudes Spatiales in France.

\bibliographystyle{plain}

\section*{Appendix}
In general, our multiwavelength data are taken from \citet{zhang12}. Therefore, we refer to the reader to that paper for details. Here, we briefly discuss the SED variability for individual sources. Unless otherwise stated, the LAT data are from the 2FGL (\citealt{ackermann11}), which are integrated between 2008 August and 2010 August, and the 1FHL (Ackermann et al., in preparation), which are integrated between 2008 August and 2011 August.

\begin{itemize}
\item Mrk~421: The data for the SED of this source are entirely simultaneous, and little variability was detected, although a few small LAT flares are evident (\citealt{abdo11d}). There does not seem to be correlated variability between the X-rays and LAT $\gamma$-rays.

\item Mrk~501: The SED is from 2006, taken from \citet{anderhub09}, and is not simultaneous with the LAT data. Variability was of the order of a factor of two in VHE $\gamma$-rays, X-rays, and optical. Variability did not appear to be correlated, although the observations are quite sparse.

\item 1ES~2344+514: Most of the multi-wavelength data are taken from 2005-2006 (\citealt{albert07a}) and are not variable, nor are they simultaneous with the LAT data. The LAT data showed minimal variability as well.

\item 1ES~1959+650: The data are from a multiwavelength campaign in 2006 where the optical and X-ray bands were in high state and showing significant variability. However, the VHE data were at the lowest fluxes ever detected for this source.

\item PKS~2005$-$489: No significant variability on timescale less than a year was found by \citet{aharonian05}. The X-ray data are from a high state of X-ray emission taken in 1998, whereas the VHE data are from 2009 when the source was in a similar X-ray state as in 1998 (\citealt{kaufmann09}).

\item W~Comae: The TeV data were taken during a strong VHE flare in 2008 (\citealt{acciari08}).

\item PKS~2155$-$304: All the data, including the LAT data are simultaneous, and variability was on the order of a factor of two in the X-rays and VHE $\gamma$-rays. The X-ray, VHE $\gamma$-ray, and optical variability appears to be approximately correlated.

\item H~1426+428: No simultaneous broadband data are found for this source. 

\item 1ES~0806+524: No significant variability was found on a timescale of months (\citealt{acciari09a}).

\item H~2356$-$309: The TeV and X-ray data were taken simultaneously in 2005 (\citealt{abramowski10}), whereas the data for the optical and other X-ray bands were taken in 2004. In the VHE, significant variations of flux were detected in a timescale of months.

\item 1ES~1218+304: The optical as well as the X-ray data were taken quasi-simultaneously in 2005. The VHE data were taken between the end of 2008 and middle of 2009. The TeV emission for this source showed day-scale variability.

\item 1ES~1101$-$232: A multiwavelength campaign for this blazar was carried out from 2004 to 2006 covering different parts of the X-ray SED showing no variability. The VHE detections were taken at the beginning of 2006.

\item 1ES~1011+496: The VHE data were taken after an optical outburst in 2007 (\citealt{albert07b}), which may indicate a correlation between the optical and VHE flux. The X-ray data were taken in 2008 May.

\item PG~1553+113: No variability for this source has been found neither in the VHE regime nor X-ray by different instruments and campaigns (\citealt{reimer08,aleksic10}).

\item 3C~66A: The simultaneous observed broadband SED is from \citet{abdo11a}. The VHE data were taken by MAGIC (\citealt{aleksic11a}) and VERITAS (\citealt{acciari09b}) in similar flux states.

\end{itemize}

\clearpage
\begin{landscape}
\begin{deluxetable}{lcccccccccccccc}
\tablecolumns{15}
\tablecaption{Synchrotron/SSC parameters of our catalog of blazars.}
\tablewidth{0pt}
\tablehead{Source & Redshift & $t_{v,min}^{(a)}$ [s] & $p_{1}^{(b)}$ & $p_{2}^{(b)}$ & $\gamma_{min}^{(c)}$ & $\gamma_{brk}^{(c)}$ & $\gamma_{max}^{(c)}$ & $R_{blob}^{(d)}$ [cm] & $\delta_{D}^{(e)}$ & $B^{(f)}$ [mG] & $L_{jet,e}^{(g)}$ [erg~s$^{-1}$] & $L_{jet,B}^{(h)}$ [erg~s$^{-1}$] & $\chi^{2}$/dof}
\startdata
Mkn~421 & 0.031 & fast & 2.25 & -- & 600 & - & $2.2\times 10^{5}$ & $1.2\times 10^{16}$ & 42 & 44 & $1.6\times 10^{44}$  & $4.1\times 10^{42}$ & 8.7/6\\
& & slow & 2.27 & -- & 1000 & - & $3.0\times 10^{5}$ & $4.9\times 10^{16}$ & 17 & 61 & $1.9\times 10^{43}$ & $9.3\times 10^{43}$ & 28.3/6\\

Mkn~501 & 0.034 & fast & 2.4 & 3.4 & 1000 & $3.2\times 10^{6}$ & $10^{4}$ & $1.7\times 10^{16}$ & 58.5 & 9.4 & $3.8\times 10^{44}$ & $6.5\times 10^{41}$ & 6.0/6\\
& & slow & 2.3 & -- & 1000 & -- & $10^{6}$ & $1.1\times 10^{17}$ & 38 & 2 & $9.8\times 10^{44}$ & $5.3\times 10^{41}$ & 2.5/6\\

1ES~2344+514 & 0.044 & fast & 2.4 & -- & $7.8\times 10^{3}$ & -- & $1.2\times 10^{5}$ & $5.6\times 10^{16}$ & 20 & 140 & $1.9\times 10^{42}$ & $1.0\times 10^{43}$ & 30.4/5\\
& & slow & 2.4 & -- & $1.9\times 10^{4}$ & -- & $5.7\times 10^{5}$ & $3.5\times 10^{16}$ & 12 & 23 & $7.2\times 10^{41}$ & $4.3\times 10^{43}$ & 56.3/5\\

1ES~1959+650 & 0.048 & fast & 2.2 & 2.3 & 1 & $3.2\times 10^{3}$ & $8.9\times 10^{4}$ & $4.4\times 10^{15}$ & 15.4 & 1800 & $1.1\times 10^{44}$ & $3.1\times 10^{43}$ & 31.4/6\\
& & slow & 2.2 & 2.3 & 1 & $6.6\times 10^{3}$ & $1.5\times 10^{5}$ & $2.1\times 10^{16}$ & 7.4 & 1400 & $3.4\times 10^{44}$ & $3.2\times 10^{43}$ & 31.2/6\\

PKS~2005$-$489 & 0.071 & fast & 3 & -- & $4.6\times 10^{3}$ & -- & $1.7\times 10^{6}$ & $4.3\times 10^{16}$ & 150 & 7.1 & $1.7\times 10^{43}$ & $5.6\times 10^{44}$ & 1.3/4\\
& & slow & 3 & -- & $6.7\times 10^{3}$ & -- & $3.3\times 10^{6}$ & $1.8\times 10^{17}$ & 64 & 7.7 & $5.9\times 10^{43}$ & $3.8\times 10^{44}$ & 0.6/4\\

W~Comae & 0.102 & fast & 1.5 & 4.0 & 100 & $1.1\times 10^{4}$ & $1.0\times 10^{7}$ & $2.4\times 10^{16}$ & 88 & 6.5 & $1.5\times 10^{42}$ & $2.3\times 10^{45}$ & 20.5/4\\
& & slow & 1.5 & 4.0 & 100 & $2.0\times 10^{4}$ & $1.0\times 10^{7}$ & $1.4\times 10^{17}$ & 53 & 3.2 & $4.6\times 10^{42}$ & $2.5\times 10^{45}$ & 6.0/4\\

PKS~2155$-$304 & 0.116 & fast & 2.2 & 3.4 & 100 & $2.7\times 10^{4}$ & $6.4\times 10^{5}$ & $3.2\times 10^{16}$ & 118 & 10 & $1.1\times 10^{43}$ & $4.8\times 10^{45}$ & 2.2/5\\
& & slow & 2.2 & 3.4 & 100 & $3.6\times 10^{4}$ & $8.9\times 10^{5}$ & $1.3\times 10^{17}$ & 49.9 & 13 & $5.5\times 10^{43}$ & $3.2\times 10^{45}$ & 0.5/5\\

H~1426+428 & 0.129 & fast & 2 & 3 & 1 & $7.0\times 10^{4}$ & $3\times 10^{7}$ & $5.9\times 10^{16}$ & 223 & 0.39 & $2.0\times 10^{41}$ & $1.1\times 10^{46}$ & 8.0/3\\
& & slow & 2 & 2.9 & 1 & $4.9\times 10^{4}$ & $3\times 10^{7}$ & $2.3\times 10^{17}$ & 86.8 & 0.58 & $1.0\times 10^{42}$ & $8.1\times 10^{45}$ & 11.4/3\\

1ES~0806+524 & 0.138 & fast & 1.7 & 3.1 & 1 & $1.9\times 10^{4}$ & $1.7\times 10^{5}$ & $8.6\times 10^{15}$ & 23 & 110 & $1.3\times 10^{44}$ & $7.0\times 10^{42}$ & 48.2/5\\
& & slow & 1.7 & 3.1 & 1 & $4.6\times 10^{4}$ & $5.1\times 10^{5}$ & $6.1\times 10^{16}$ & 32 & 21 & $3.6\times 10^{44}$ & $6.8\times 10^{42}$ & 31.0/5\\

H~2356$-$309 & 0.165 & fast & 2.3 & -- & 10 & -- & $9.1\times 10^{4}$ & $4.0\times 10^{15}$ & 15.5 & 1200 & $4.2\times 10^{43}$ & $3.2\times 10^{43}$ & 1.5/3\\ 
& & slow & 2 & -- & 10 & -- & $1.1\times 10^{5}$ & $1.9\times 10^{16}$ & 7.2 & 840 & $9.5\times 10^{43}$ & $1.5\times 10^{43}$ & 2.0/3\\

1ES~1218+304 & 0.182 & fast & 2.2 & -- & 1 & -- & $3.8\times 10^{5}$ & $1.8\times 10^{16}$ & 72 & 9.1 & $1.1\times 10^{42}$ & $5.8\times 10^{45}$ & 5.2/5\\
& & slow & 2.2 & -- & 1 & -- & $1.7\times 10^{6}$ & $2.2\times 10^{17}$ & 86 & 0.38 & $3.7\times 10^{41}$ & $6.3\times 10^{46}$ & 4.4/5\\

1ES~1101$-$232 & 0.186 & fast & 2 & -- & 1000 & -- & $3.0\times 10^{6}$ & $1.3\times 10^{17}$ & 500 & 0.04 & $3.8\times 10^{46}$ & $4.1\times 10^{40}$ & 1.8/1\\
& & slow & 2 & -- & 1000 & -- & $2.6\times 10^{6}$ & $3.6\times 10^{17}$ & 140 & 0.17 & $8.8\times 10^{45}$ & $5.7\times 10^{41}$ & 1.9/1\\

1ES~1011+496 & 0.212 & fast & 2 & 5.4 & 1 & $2.7\times 10^{4}$ & $1.0\times 10^{8}$ & $3.6\times 10^{15}$ & 14.6 & 6900 & $1.0\times 10^{44}$ & $2.4\times 10^{43}$ & 6.4/5\\ 
& & slow & 2 & 5.4 & 1 & $5.6\times 10^{4}$ & $1.0\times 10^{8}$ & $2.2\times 10^{16}$ & 9.1 & 2300 & $1.7\times 10^{45}$ & $4.7\times 10^{43}$ & 7.2/5\\ 

PG~1553+113 & 0.395 & fast & 2 & 3.8 & 1 & $3.9\times 10^{4}$ & $2.6\times 10^{6}$ & $4.8\times 10^{16}$ & 200 & 6.3 & $2.3\times 10^{46}$ & $3.4\times 10^{43}$ & 13.7/5\\
& 0.395 & slow & 2 & 3.8 & 1 & $6.6\times 10^{4}$ & $4.4\times 10^{6}$ & $2.4\times 10^{17}$ & 110 & 4.6 & $2.3\times 10^{46}$ & $1.1\times 10^{44}$ & 4.2/5\\
& 0.58 & fast & 2 & 3.8 & 1 & $3.3\times 10^{4}$ & $2.2\times 10^{6}$ & $4.6\times 10^{16}$ & 240 & 9.2 & $2.8\times 10^{46}$ & $7.8\times 10^{43}$ & 21.4/5\\
& 0.58 & slow & 2 & 3.8 & 1 & $7.4\times 10^{4}$ & $5.0\times 10^{6}$ & $2.9\times 10^{17}$ & 150 & 0.03 & $6.5\times 10^{47}$ & $5.2\times 10^{46}$ & 3.8/5\\

3C~66A & 0.444 & fast &  2.5 & 3.3 & 300 & $4.0\times 10^{4}$ & $1.3\times 10^{5}$ & $6.1\times 10^{16}$ & 290 & 3.8 & $2.3\times 10^{47}$ & $3.4\times 10^{43}$ & 24.1/5\\
& & slow & 2.5 & 3.3 & 300 & $4.2\times 10^{4}$ & $1.8\times 10^{5}$ & $2.5\times 10^{17}$ & 120 & 5.6 & $1.5\times 10^{47}$ & $2.0\times 10^{44}$ & 24.9/5\\
\enddata
\tablenotetext{\,}{$\bf{Note.-}$ The 15 blazars in our catalog are listed with their position in the sky in equatorial J2000 coordinates and estimated redshifts. The two best-fitting sets of parameters of the quasi-simultaneous multiwavelength data to a one-zone synchrotron/synchrotron self-Compton model are listed as well. The two models for each blazar are mainly characterized by different minimum variability timescale (see column $(a)$; $t_{v,min}=10^{4}$~s for the fast model and $t_{v,min}=10^{5}$~s for the slow model). These two fits bracket the expected VHE flux derived from the same set of lower-energy multiwavelength data. $(a)$ Minimum variability timescale, $(b)$ Electron-distribution index, $(c)$ Minimum, maximum, and break electron Lorentz factors, $(d)$ Blob radius, $(e)$ Doppler factor, $(f)$ Magnetic field strength, $(g)$ Electric-field power in the jets, $(h)$ Magnetic-field power in the jets, $(i)$ $\chi^{2}$ divided by the degrees of freedom. We note that the fit of the blazar 1ES~1101-232 has only one degree of freedom.}
\label{tab1}
\end{deluxetable}
\clearpage
\end{landscape}

\clearpage
\begin{landscape}
\begin{deluxetable}{lccccc}
\tablecolumns{5}
\tablecaption{The measured and predicted cosmic $\gamma$-ray horizon.}
\tablewidth{0pt}
\tablehead{Source & Redshift & $E_{0}\pm \Delta E_{0}~({\rm fast}/{\rm slow})^{(a)}$~[TeV] & $E_{0}\pm (\Delta E_{0})_{stat}\pm (\Delta E_{0})_{sys}^{(b)}$~[TeV] & $E_{D11}\pm \Delta E_{D11}^{(c)}$~[TeV] & IACT reference}
\startdata
Mkn~421 & 0.031 & $10.42\pm 7.72$/$11.91\pm 8.79$ & $11.14_{-8.44}^{+9.56}\pm 2.23$ & $9.72_{-3.17}^{+1.85}$ & \citet{abdo11d}\\
Mkn~501 & 0.034 & $11.85\pm 16.84$/$2.28\pm 1.02$ & $5.20_{-3.94}^{+23.49}\pm 1.04$ & $8.75_{-3.31}^{+1.68}$ & \citet{acciari11}\\
1ES~2344+514 & 0.044 & None/$6.65\pm 21.49$ & None & $6.01_{-3.23}^{+1.20}$ & \citet{albert07a}\\
1ES~1959+650 & 0.048 & None/None & None & $5.12_{-2.99}^{+1.02}$ & \citet{tagliaferri08}\\
PKS~2005$-$489 & 0.071 & $1.99\pm 0.26$/$2.09\pm 0.25$ & $2.04_{-0.31}^{+0.30}\pm 0.41$ & $1.83_{-1.06}^{+0.34}$ & \citet{kaufmann09}\\
W~Comae & 0.102 & None/None & None & $0.90_{-0.18}^{+0.09}$ & \citet{acciari08}\\
PKS~2155$-$304 & 0.116 & $0.77\pm 0.17$/$0.88\pm 0.05$ & $0.82_{-0.22}^{+0.11}\pm 0.16$ & $0.77_{-0.13}^{+0.07}$ & \citet{aharonian09}\\
H~1426+428 & 0.129 & $6.23\pm 7.64$/$13.24\pm 16.81$ &  None & $0.68_{-0.11}^{+0.06}$ & \citet{aharonian02}\\
1ES~0806+524 & 0.138 & $0.35\pm 0.04$/$0.85\pm 0.01$ & $0.55_{-0.24}^{+0.31}\pm 0.11$ & $0.64_{-0.10}^{+0.05}$ & \citet{acciari09a}\\
H~2356-309 & 0.165 & None/None & None & $0.54_{-0.07}^{+0.04}$ & \citet{abramowski10}\\
1ES~1218+304 & 0.182 & $0.58\pm 0.02$/$0.46\pm 0.02$ & $0.52_{-0.08}^{+0.08}\pm 0.10$ & $0.49_{-0.06}^{+0.04}$ & \citet{acciari10}\\
1ES~1101$-$232 & 0.186 & $0.41\pm 0.02$/$0.39\pm 0.01$ & $0.40_{-0.02}^{+0.03}\pm 0.08$ & $0.48_{-0.06}^{+0.04}$ & \citet{aharonian06}\\
1ES~1011+496 & 0.212 & None/None & None & $0.43_{-0.05}^{+0.03}$ & \citet{albert07b}\\
3C~66A & 0.444 & $0.29\pm 0.02$/$0.31\pm 0.02$ & $0.30_{-0.03}^{+0.03}\pm 0.06$ & $0.23_{-0.02}^{+0.02}$ & \citet{acciari09b,aleksic11a}\\
PG~1553+113 & 0.500$^{+0.080}_{-0.105}$ & $0.24\pm 0.01$/$0.23\pm 0.02$ & $0.23_{-0.03}^{+0.05}\pm 0.05$ & $0.21_{-0.02}^{+0.02}$ & \citet{aleksic10}\\
\enddata
\tablenotetext{\,}{$\bf{Note.-}$ The 15 blazars in our catalog are listed with their estimated redshifts. The energy $E_{0}$ is shown for the two different variability timescales ($t_{v,min}=10^{4}$~s for the fast model and $t_{v,min}=10^{5}$~s for the slow model) and its combined value as described in the text in the columns $(a)$ and $(b)$, respectively. The combined $E_{0}$ is given with its statistical and systematic uncertainties (see text for details). The $E_{0}$ estimated from the EBL model discussed in \citet{dominguez11a} ($E_{D11}\pm \Delta E_{D11}$) is given as well in column $(c)$. None means that our methodology output no solution for the $E_{0}$.}
\label{tab2}
\end{deluxetable}
\clearpage
\end{landscape}

\label{lastpage}
\end{document}